\documentclass[iop]{emulateapj}
\usepackage{apjfonts}
\usepackage{natbib}
\usepackage{subfigure}
\usepackage{mathrsfs}
\usepackage{booktabs}
\usepackage{amsmath}

\newcommand{\xmm}{\hbox{\it XMM-Newton\/}}
\newcommand{\chandra}{{\it Chandra\/}}
\newcommand{\swift}{{\it Swift\/}}
\newcommand{\nustar}{{\it NuSTAR\/}}

\newcommand{\xray}{\hbox{\it \rm X-ray\/}}

\begin{document}
\title{SDSS J075101.42+291419.1: A Super-Eddington Accreting Quasar with Extreme X-ray Variability} 
\author{
Hezhen~Liu\altaffilmark{1,2,3},
B.~Luo\altaffilmark{1,2,3},
W. N. Brandt\altaffilmark{4,5,6},
Michael S. Brotherton\altaffilmark{7},
Pu Du\altaffilmark{8},
S.~C.~Gallagher\altaffilmark{9},
Chen Hu\altaffilmark{8},
Ohad~Shemmer\altaffilmark{10},
Jian-Min Wang\altaffilmark{8, 11, 12}
}
\altaffiltext{1}{School of Astronomy and Space Science, Nanjing University, Nanjing, Jiangsu 210093, China}
\altaffiltext{2}{Key Laboratory of Modern Astronomy and Astrophysics (Nanjing University), Ministry of Education, Nanjing 210093, China}
\altaffiltext{3}{Collaborative Innovation Center of Modern Astronomy and Space Exploration, Nanjing 210093, China}
\altaffiltext{4}{Department of Astronomy \& Astrophysics, 525 Davey Lab,
The Pennsylvania State University, University Park, PA 16802, USA}

\altaffiltext{5}{Institute for Gravitation and the Cosmos,
The Pennsylvania State University, University Park, PA 16802, USA}

\altaffiltext{6}{Department of Physics, 104 Davey Lab,
The Pennsylvania State University, University Park, PA 16802, USA}
\altaffiltext{7}{Department of Physics and Astronomy, University of Wyoming, Laramie, WY 82071, USA}
\altaffiltext{8}{Key Laboratory for Particle Astrophysics, Institute of High Energy Physics,
Chinese Academy of Sciences, 19B Yuquan Road, Beijing 100049, China}
\altaffiltext{9}{
	Department of Physics \& Astronomy and Centre for Planetary and Space Exploration, The University of Western
  Ontario, London, ON, N6A 3K7, Canada}
\altaffiltext{10}{
	Department of Physics, University of North Texas, Denton, TX 76203, USA} 	
\altaffiltext{11}{National Astronomical Observatories of China, Chinese Academy of Sciences,
20A Datun Road, Beijing 100020, China}
\altaffiltext{12}{School of Astronomy and Space Science, University of Chinese Academy of Sciences,
19A Yuquan Road, Beijing 100049, China}

\begin{abstract}
We report the discovery of extreme \xray\
 variability in a type 1 quasar: SDSS J075101.42+291419.1.
It has a black hole mass of $1.6\times 10^7~\rm M_\odot$
 measured from reverberation mapping (RM), and the black
 hole is accreting with a \hbox{super-Eddington} accretion
 rate. 
 Its \xmm\ observation in 2015 May reveals a flux drop by a factor of $\sim 22$ 
 with respect to the \swift\ observation in 2013 May when it showed a typical 
 level of \xray\ emission relative to its UV/optical emission.
The lack of correlated UV variability results in a steep
\hbox{X-ray-to-optical} power-law slope ($\alpha_{\rm OX}$)
of $-1.97$ in the low \xray\ flux state, corresponding to an \xray\
 weakness factor of 36.2 at rest-frame 2~keV relative to 
 its UV/optical luminosity. The mild UV/optical continuum and emission-line variability also suggest that the accretion rate did not change significantly.
A single power-law model modified by Galactic absorption
 describes well the 0.3--10~keV spectra of the \xray\
 observations in general.
The spectral fitting reveals steep spectral shapes with $\Gamma\approx
3$.
We search for active galactic nuclei (AGNs) with such extreme
 \xray\ variability in the literature and find that most of
  them are narrow-line Seyfert 1 galaxies and 
  quasars with high accretion rates.
The fraction of extremely \xray\ variable 
 objects among super-Eddington accreting AGNs is estimated to be 
 $\approx 15\textrm{--}24\%$. 
We discuss two possible scenarios, disk reflection and
partial covering absorption, to explain the extreme \xray\
 variability of SDSS J$075101.42+291419.1$.
We propose a possible origin for the partial covering absorber,
which is the thick inner accretion disk and its associated outflow 
in AGNs with high accretion rates.
\end{abstract}


\section{Introduction}\label{sec:intro}
Active galactic nuclei (AGNs) are in general powerful \xray\ sources
in the Universe. 
\xray\ emission from AGNs mainly originates from a
 corona of hot electrons in the vicinity of the black hole (BH).
Thermal UV/optical photons emitted from the accretion disk are
\hbox{inverse-Compton} scattered by the hot electrons to 
\xray\ energies, forming a \hbox{power-law} continuum 
\citep[e.g.,][]{Sunyaev1980,Haardt1993}.
The detailed physics of the corona is not well understood. 
Observationally, except for radio-loud AGNs and other AGNs 
with potential \xray\ absorption, such as broad absorption 
line (BAL) quasars, the \xray\ emission of type 1 AGNs is 
closely related to the UV/optical emission from the accretion disk.
This is generally quantified as a tight correlation between the 
\hbox{\xray-to-optical} \hbox{power-law} slope parameter  
 ($\alpha_{\rm OX}$\footnote{$\alpha_{\rm OX}=
0.384{\rm log}(f_{\rm 2~keV}/f_{\rm 2500~{\textup{\AA}}})$,
where ${\sc f_{\rm 2~keV}}$ and $f_{\rm 2500~{\textup{\AA}}}$ are
the \hbox{rest-frame} 2~keV and $2500~{\textup{\AA}}$ flux densities.
It is an \xray\ loudness parameter that is used to compare
the UV/optical and \xray\ luminosities.})
and the $2500~{\textup{\AA}}$ monochromatic luminosity 
\citep[e.g.,][]{Strateva2005,Steffen2006,Just2007,Grupe2010,Lusso2010}.
 This relation is often used to assess the level of \xray\ weakness
 for AGNs. When the observed $\alpha_{\rm OX}$ value 
 of an AGN is lower than that expected from the
 \hbox{$\alpha_{\rm OX}$--$L_{\rm 2500~{\textup{\AA}}}$} relation
 ($\Delta\alpha_{\rm OX}=\alpha_{\rm OX}-\alpha_{\rm OX,exp}<0$, 
 where more negative values indicate more extreme \xray\ weakness 
 relative to the $\rm 2500~{\textup{\AA}}$ monochromatic luminosity), 
 it is \xray\ weak by a linear factor of  
 $f_{\rm weak}=10^{-\Delta\alpha_{\rm OX}/0.384}$.
   
 X-ray luminosity variability is a characteristic property of AGNs,
which is generally related to instabilities of
 the corona or fluctuations of the accretion flow \citep[e.g.,][]
 {Nandra2001,McHardy2006,MacLeod2010,Yang2016,Zheng2017}.
The intrinsic \xray\ variability of AGNs contributes partially to
the dispersion of the
\hbox{$\alpha_{\rm OX}$--$L_{\rm 2500~{\textup{\AA}}}$} relation
 \citep[e.g.,][]{Vagnetti2010,Gibson2012,Vagnetti2013,Chiaraluce2018}.
Compared to AGN variability at longer wavelengths, AGN \xray\ 
variability often displays larger amplitudes on timescales of years
down to minutes \cite[e.g.,][]{Ulrich1997,Peterson2001}.
The typical long-term \xray\ flux variability amplitude of AGNs is 
$\sim 20\textrm{--}50\%$, and seldom exceeds $200\%$ \citep[e.g.,][] 
{Yuan1998,Grupe2001,Paolillo2004,Mateos2007,Vagnetti2011,Gibson2012,
Soldi2014,Yang2016,Middei2017,Maughan2018}. 
There are several types of AGNs that may have excess \xray\
 variability compared to typical AGNs.
Radio-loud AGNs are known to have \xray\ emission contributed from 
the jets \citep[e.g.,][]
{Worrall1987,Wilkes1987,Worrall2006,Miller2011,Komossa2018},
and they often show strong \xray\ variability related to the jets
 \citep[e.g.,][]{Gliozzi2007,Carnerero2017,Zhu2018}.
Strong \xray\ variability has also been observed in several BAL
quasars, likely due to \xray\ absorption variability 
\citep[e.g.,][]{Gallagher2004,Saez2012,Kaastra2014,Mehdipour2017}. 
Recently, a small number of \hbox{"changing-look"} quasars 
have been discovered \cite[e.g.,][]
{LaMassa2015,Parker2016,Mathur2018,Oknyansky2019}.
Such an AGN likely has different amounts of \xray\ emission in
different states, although its $\alpha_{\rm OX}$ values are not
necessarily abnormal as the \xray\ and UV/optical emission might 
vary in a coordinated manner.
We do not consider these types of AGNs in our following 
discussion of AGN \xray\ variability.

Besides radio-loud AGNs, BAL quasars, and \hbox{changing-look}
 quasars, only a small number of extremely \xray\ variable AGNs 
 have been reported in the past two decades; in this study 
we consider extremely \xray\ variable AGNs as those  
 varying in X-rays by factors of larger than 10 
 ($\ga 2.5\sigma$ deviation from the expected 
 \hbox{$\alpha_{\rm OX}$--$L_{\rm 2500~{\textup{\AA}}}$} relation;
 Table~5 of \citealt{Steffen2006}).
These AGNs all vary between the \xray\ normal and weak states, 
with \xray\ weakness factors of $f_{\rm weak}>10$ 
($\Delta\alpha_{\rm OX} <-0.384$) in the low state.
 Most of these objects are narrow-line Seyfert 1 galaxies
  (NLS1s), which are a well-studied group of AGNs that generally
  have small BH masses ($\la 10^7~\rm M_\odot$) and large
  Eddington ratios ($\lambda_{\rm Edd}\ga 0.1$; e.g., \citealt{Boller1996,Leighly1999a,
  Leighly1999b,Grupe2001,Grupe2004,Gallo2018_nls1,Komossa2018}
  and references therein).  
We give a few examples of this phenomenon observed in NLS1s. 
Mrk~335 fell into a historically
low \xray\ flux state in 2007 with a flux drop by a factor of 30,
and it has ever since shown persistent \xray\ variability
and occasional intense flares with variability factors of $\approx 10$
in nearly 11 years of monitoring observations
\citep[e.g.,][]{Grupe2007a,Grupe2012,Gallo2018}.
NGC~4051 was observed in an extremely dim state in 1998, 
with the \xray\ flux about 20 times fainter 
 than its historical average value 
 \citep[e.g.,][]{Guainazzi1998,Uttley1999,Peterson2000}.  
1H~$0707-495$ and IRAS~$13224-3809$ exhibit very rapid \xray\
 variability with variability amplitudes up to
 $\approx 100\textrm{--}200$ within only a few hours
 \citep[e.g.,][]{Boller2002,Fabian2009,Fabian2012,Fabian2013,
 Ponti2010,Jiang2018}.
 
At higher luminosities, only three radio-quiet non-BAL quasars 
have been found to show extreme \xray\ variability (by factors 
of larger than 10), which are PHL~1092 \citep[e.g.,][]
{Miniutti2009,Miniutti2012}, PG~$0844+349$
\citep[e.g.,][]{Gallagher2001,Gallo2011}, and PG~$1211+143$
\citep[e.g.,][]{Bachev2009}. 
In this study, we consider quasars as AGNs with the  
rest-frame $5100~\textup{\AA}$ luminosity 
larger than $10^{44}~\rm erg~s^{-1}$, which typically have 
BH masses of $\ga 10^7 \rm~M_\odot$.   
These three extremely \xray\ variable quasars are narrow-line
 (NL) type 1 quasars that possess 
narrower than typical $\rm H\beta$ lines in their optical
spectra similar to NLS1s, indicating relatively small BH 
masses and high Eddington ratios compared to typical quasars.   
Another common feature among these extremely \xray\ variable
 NLS1s and NL type 1 quasars is that there is no coordinated 
 UV/optical continuum or emission-line variability with their 
 \xray\ variability, suggesting that the accretion 
 rates of these objects did not change significantly
 \cite[e.g.,][]{Bachev2009,Gallo2011,Grupe2012,Miniutti2012,
 Robertson2015,Buisson2018}.
Most of the extremely \xray\ variable AGNs mentioned 
above have been discussed in early studies of $\alpha_{\rm OX}$ 
variability in Seyfert 1 galaxies, and it has been suggested 
that extremely variable $\alpha_{\rm OX}$ values are 
preferentially observed in NLS1s \citep[e.g.,][]
{Gallo2006,Vasudevan2011}.

The rarity of extremely \xray\ variable AGNs discovered so far
may be related to the limited numbers of \xray\ observations 
available for individual objects, and we probably have not 
found the extremely \xray\ weak states for most of the NLS1s
 and NL type 1 quasars. Using the published data of two
  extremely \xray\ variable AGNs, Mrk~335 and PHL~1092, we 
  roughly estimated that 
 the duty cycles of their extremely \xray\ weak states 
 ($f_{\rm weak}>10$, $\Delta\alpha_{\rm OX}<-0.384$) are $30\%$ 
 and $60\%$, respectively (see the analysis in
 Section~\ref{subsec:frac} below). If extreme \xray\ variability is 
 a common feature for NLS1s and NL type 1 quasars, with a similar 
  $\approx 30\textrm{--}60 \%$ duty cycle for the extremely 
  \xray\ weak state, we would expect that among a large sample 
  of these AGNs, $\approx 30\textrm{--}60 \%$ will deviate
 significantly from the \hbox{$\alpha_{\rm OX}$--$L_{\rm
 2500~{\textup{\AA}}}$} relation (with $\Delta\alpha_{\rm OX}
 <-0.384$). However, such large fractions of \xray\ weak 
 outliers have not been found 
 in previous studies of the \hbox{$\alpha_{\rm OX}$--$L_{\rm
  2500~{\textup{\AA}}}$} relation \citep[e.g.,][]
 {Grupe2010,Vasudevan2011,Gibson2012}.   
  The rarity of extreme \xray\ variable AGNs is likely
  intrinsic, and they only constitute a small fraction of
  NLS1 or NL type 1 quasar population 
  (see Section~\ref{subsec:frac} below
  for detailed discussion).

 Here we report the discovery of extreme
X-ray variability in another quasar, SDSS J$075101.42+291419.1$ 
(hereafter SDSS J$0751+2914$). The source has a redshift of 
$z=0.1208$, and it is a radio-quiet
(radio loudness $R=1.29$)\footnote{The radio loudness parameter 
is defined as $R=f_{\rm 6 ~cm}/f_{\rm 2500~\textup{\AA}}$,
 where $f_{\rm 6~cm}$ and $f_{\rm 2500~\textup{\AA}}$ are flux densities
 at 6~cm and $2500~\textup{\AA}$, respectively \cite[e.g.,][]{Jiang2007}. 
 The value is adopted from \cite{Shen2011}.}
 quasar that possesses spectroscopic characteristics similar to NLS1s;
 e.g., a relatively narrow $\rm H\beta$ line ($\rm FWHM=1679~km~s^{-1}$),
 strong optical \ion{Fe}{2} emission, and weak [\ion{O}{3}] lines.
It is one of the targets in the reverberation mapping (RM)
campaign targeting super-Eddington accreting massive black holes
\citep[SEAMBHs;][]{Du2015,Du2016,Du2018}.
Its virial BH mass constrained from the $\rm H\beta$ RM is
$1.6\times 10^7~\rm M_\odot$, and its dimensionless mass accretion
 rate derived from the standard thin disk model 
 \citep{Shakura1973} is $\dot{\rm \mathscr{M}}=\dot{M}c^2/
 {L_{\rm Edd}= 20.1 (\ell_{44}}/{\cos i})^{3/2}(M_{\rm BH}
 /10^7M_\odot)^{-2}=28.2$, where $\dot{M}$ is the mass accretion
  rate,  $L_{\rm Edd}=1.5\times 10^{38}(M_{\rm BH}/M_\odot)~\rm 
  erg~s^{-1}$ is the Eddington luminosity for solar 
   composition gas, $\ell_{44} = L_{5100~\textup{\AA}}/ 10^{44}~\rm 
erg~s^{-1}$ is the rest-frame $5100~\textup{\AA}$ luminosity in units
 of $10^{44}~\rm erg~s^{-1}$, $M_{\rm BH}$ is the BH mass, and {\it i} 
  (adopted as cos $ {\it i} =0.75$) is the inclination angle of the 
  disk \citep[e.g.,][]{Du2015,Du2018}. The Eddington ratio and 
  $\dot{\rm \mathscr{M}}$ follow the relation: $\lambda_{\rm Edd}=
  L_{\rm Bol}/L_{\rm Edd}=\eta\dot{\rm \mathscr{M}}$, 
  where $L_{\rm Bol}$ is the disk bolometric luminosity, 
  and $\eta$ is the \hbox{mass-to-radiation} conversion
  efficiency. According to the criterion of $\dot{\rm \mathscr{M}}>3$ 
  used for identifying SEAMBH candidates \citep{Wang2014a,Du2015}, 
  SDSS J$0751+2914$ can be classified as a SEAMBH candidate.  
 We caution that there could be substantial systematic  
  uncertainties on the the measured BH mass and accretion rate of 
  SDSS J$0751+2914$, because for an AGN with a high accretion rate, 
  the broad line region (BLR) may no longer be virialized 
  \citep[e.g.,][]
  {Marconi2008,Marconi2009,Netzer2010,Krause2011,Pancoast2014,Li2018}
  and the accretion disk may deviate from the standard thin disk
   \citep[e.g.,][]{Abramowicz1980,Abramowicz1988,Wang2003,Ohsuga2011,
 Wang2014b, Jiang2016,Jiang2017}.   
SDSS~J$0751+2914$ has archival \swift\ \citep{Gehrels2004} and \xmm\
\citep{Jansen2001} \xray\ observations,
for which simultaneous UV/optical photometric observations by
the same satellites are available. 
We serendipitously discovered its extreme \xray\ variability
during systematic analyses of the \xray\ properties for SEAMBHs 
(H. Liu et al, in preparation).

The paper is organized as follows. In Section~\ref{sec:data},
we present the mutiwavelength observations and describe the data
 analysis processes. The main results are presented in
   Section~\ref{sec:result}. In Section~\ref{sec:discuss},
   we describe the common properties of extremely \xray\ variable
   AGNs, estimate the occurrence rate of extreme \xray\ variability 
   among AGNs with high accretion rates, and 
   discuss possible physical mechanisms for extreme
   \xray\ variability in AGNs.
   We summarize and present future prospects in Section~\ref{sec:sum}.
   Throughout this paper,
we use J2000 coordinates and a cosmology with
$H_0=67.4$~km~s$^{-1}$~Mpc$^{-1}$, $\Omega_{\rm M}=0.315$,
and $\Omega_{\Lambda}=0.686$ \citep{Planck2018}.

\section{Multiwavelength Observations and Data Reduction} 
\label{sec:data}
\subsection{Swift Observations}
SDSS J$0751+2914$ has been observed by \swift\ on seven 
occasions since 2013. These observations were performed 
simultaneously with
the \xray\ Telescope (XRT; \citealt{Burrows2005}) and UV-Optical
Telescope (UVOT; \citealt{Roming2005}).
For five observations, the exposure times are less than 1~ks, and 
we did not use their XRT data since no useful constraints can be
derived.
We analyzed the XRT data of the other two observations,
which were performed on 2013 May 27 and 2014 September 4, respectively. 
The observation log is given in Table~\ref{tbl-obs}.
The XRT was operated in Photon Counting (PC) mode \citep{Hill2004} and the data were reduced with the task {\it xrtpipeline} version 0.13.2, which is included in the HEASOFT package 6.17.
The spectral extraction was performed using the task {\it xselect} version 2.4.
Source photons were selected from a circular region centered on the optical position of SDSS J$0751+2914$ with a 47\arcsec\ radius.
The corresponding background photons were extracted from a nearby source-free circular region with a 100\arcsec\ radius.
The ancillary response function files (arfs) were generated by {\it xrtmkarf}, and the standard photon redistribution matrix files (rmfs) were obtained from the CALDB.
There are 176 and 34 photons in the \hbox{0.3--10~keV} band in the spectra of the 2013 and 2014 observations, respectively.
Considering the small numbers of counts,
we grouped the spectra using {\it grppha} such that each bin contains at least 1 photon count.

Each of the seven UVOT observations was performed using only one filter (U, UVW1, or UVM2),
which is reported in Table~\ref{tbl-xbasic}.
After aspect correction, the exposures were co-added for each segment in each filter using the task {\it uvotimsum}.
Source counts were extracted from a circular region with a 5\arcsec\ radius centered on the source position determined by the task {\it uvotdetect}, and the background counts were extracted from a nearby source-free circular region with a radius of 20\arcsec.
The magnitudes and fluxes in these UVOT bands were then computed using the task {\it uvotsource}.
These data were corrected for Galactic extinction at the source position ($E_{\rm B-V}=0.042$; \citealt{Schlegel1998}) following the dereddening approach of \cite{Cardelli1989} and \cite{Donnell1994}.

\subsection{XMM-Newton Observation}
SDSS J0751+2914 was observed with \xmm\ for $\sim 11$~ks on 2015 May
4 using simultaneously the European Photon Imaging Camera (EPIC) PN
 \citep{Struder2001} and MOS \citep{Turner2001} detectors,
 and the Optical Monitor \citep[OM;][]{Mason2001}.
The observation information is reported in Table~\ref{tbl-obs}.
This observation was presented in \cite{Castell2017}, while
we reprocessed the observational data to make comparisons to the 
\swift\ observations. The EPIC observations were operated in Full 
Window mode.
The data were processed using the \xmm\ Science Analysis System
(SAS v.16.0.0) and the latest calibration files.
We only used the EPIC PN \xray\ data, which were reduced with the
task {\it epproc}. Only single and double events were selected,
and bad pixels and high background flares were filtered from the
 calibrated event lists based on the standard selection criteria,
 which resulted in a final cleaned exposure time of $5.3$~ks.
We extracted the source spectrum using a circular region with a radius
of 35\arcsec centered on the source position determined by the task
{\it edetect-chain}. The background spectrum was extracted from a
 nearby source-free circular region of the same size
 in the same CCD chip. Spectral response files were generated using 
 the tasks {\it rmfgen} and {\it arfgen}.
The source spectrum contains 295 photons in the 0.3--10~keV energy
 band. We grouped the spectrum using the task {\it specgroup} with 
 a minimum of one photon count per energy bin.

The OM has similar filters to those of the \swift\ UVOT, 
although the effective wavelengths of these filters are somewhat
 different (see Table~\ref{tbl-obs}). The OM observation was reduced
with the task {\it omchian}, which generated five exposures for 
three filters (U, UVW1, and UVM2).
The photometric results of every exposure are recorded in SWSRLI
files, and we extracted the magnitudes and fluxes of our target from
these files. We adopted the mean magnitudes and fluxes of all the
exposure segments for each filter,
which were then corrected for Galactic extinction.

\subsection{Lijiang Observations}
SDSS $\rm J0751+2914$ was observed repeatedly with the
Lijiang 2.4~m telescope at the Yunnan Observatories of the 
Chinese Academy of Sciences during 2013 November--2014 May
and 2016 October--2017 June \citep{Du2015,Du2018}.
It was also observed with the Lijiang telescope simultaneously with
 the \xmm\ observation on 2015 May 4.
The details of the observations and the data reduction were reported 
in \cite{Du2014,Du2015,Du2018}.
The light curves of the $5100~{\textup{\AA}}$ continuum flux density 
and the $\rm H\beta$ emission-line flux in the two RM monitoring 
periods are presented in Section~\ref{subsec:opt} below.
We gathered three Lijiang spectra, for which the observation dates 
are closest to those of the \xray\ observations.
The first spectrum was observed on 2013 November 13, 170 days after
 the 2013 \swift\ observation. The second spectrum was observed on 
 2014 May 17, 110 days before the 2014 \swift\ observation.
Another spectrum is the one observed simultaneously with the 2015 
\xmm\ observation.

\begin{deluxetable*}{lcccc}
\tabletypesize{\scriptsize}
\tablewidth{0pt}
\tablecaption{Observation Log \label{tbl-obs}
}
\tablehead{
\colhead{}  &
\colhead{}  &
\colhead{Observation} &
\colhead{Exposure Time}   &
\colhead{Bandpass or}  \\
\colhead{Observatory and Instrument}   &
\colhead{Date}   &
\colhead{ID}   &
\colhead{(s)}   &
\colhead{Effective Wavelength}
}
\startdata
\swift\ XRT (PC mode)& 2013--05--27 & 00039549001&4110 & 0.3--10~keV \\
\swift\ UVOT (UVW1)  & 2013--05--27 & 00039549001   &4065 & 2600 \AA   
\\ 
\swift\ UVOT (UVW2)  & 2014--01--25 &00039550001& 101 & 1928 \AA 
\\
\swift\ XRT (PC mode)& 2014--09--04 &00039550002 &2015 & 0.3--10~keV\\
\swift\ UVOT (U)     & 2014--09--04  &  00039550002    &2106 & 3465 \AA   
\\
\swift\ UVOT (UVW1)    & 2015--03--18 &00039550003& 158 & 2600 \AA 
\\
\swift\ UVOT (UVW2)    & 2015--05--28 &00039550004& 901 & 1928 \AA 
\\
\swift\ UVOT (UVW1)    & 2015--09--14 &00039550005&626 & 2600 \AA 
\\
\swift\ UVOT (U)       & 2016--05--20 &00039550007& 394 & 3465 \AA 
\\
\xmm\ PN               & 2015--05--04 &0761510101 & 10650 & 0.3--12~keV \\
\xmm\ MOS1             & 2015--05--04 &0761510101 & 12543 & 0.5--10~keV \\
\xmm\ MOS2             & 2015--05--04 & 0761510101 & 12649 & 0.5--10~keV \\
\xmm\ OM (U)           & 2015--05--04 & 0761510101 & 8000  & 3440 \AA \\
\xmm\ OM (UVW1)        & 2015--05--04 & 0761510101 & 8000 & 2910 \AA \\
\xmm\ OM (UVM2)        & 2015--05--04 & 0761510101 & 4000 & 2310 \AA
\enddata
\end{deluxetable*}

\section{RESULTS}\label{sec:result}

\begin{figure}
 \centering
  \includegraphics[scale=0.42]{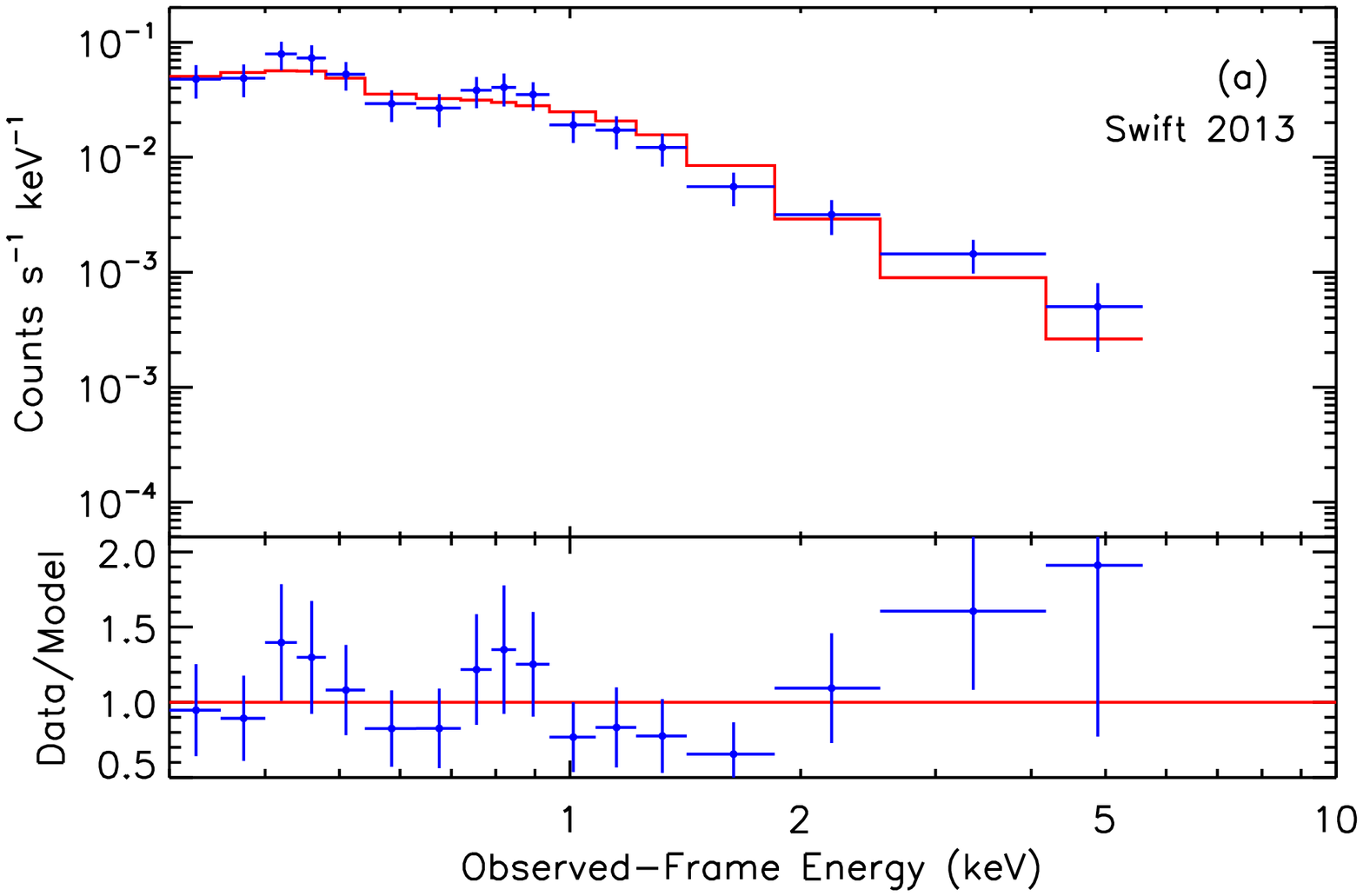}
  \includegraphics[scale=0.42]{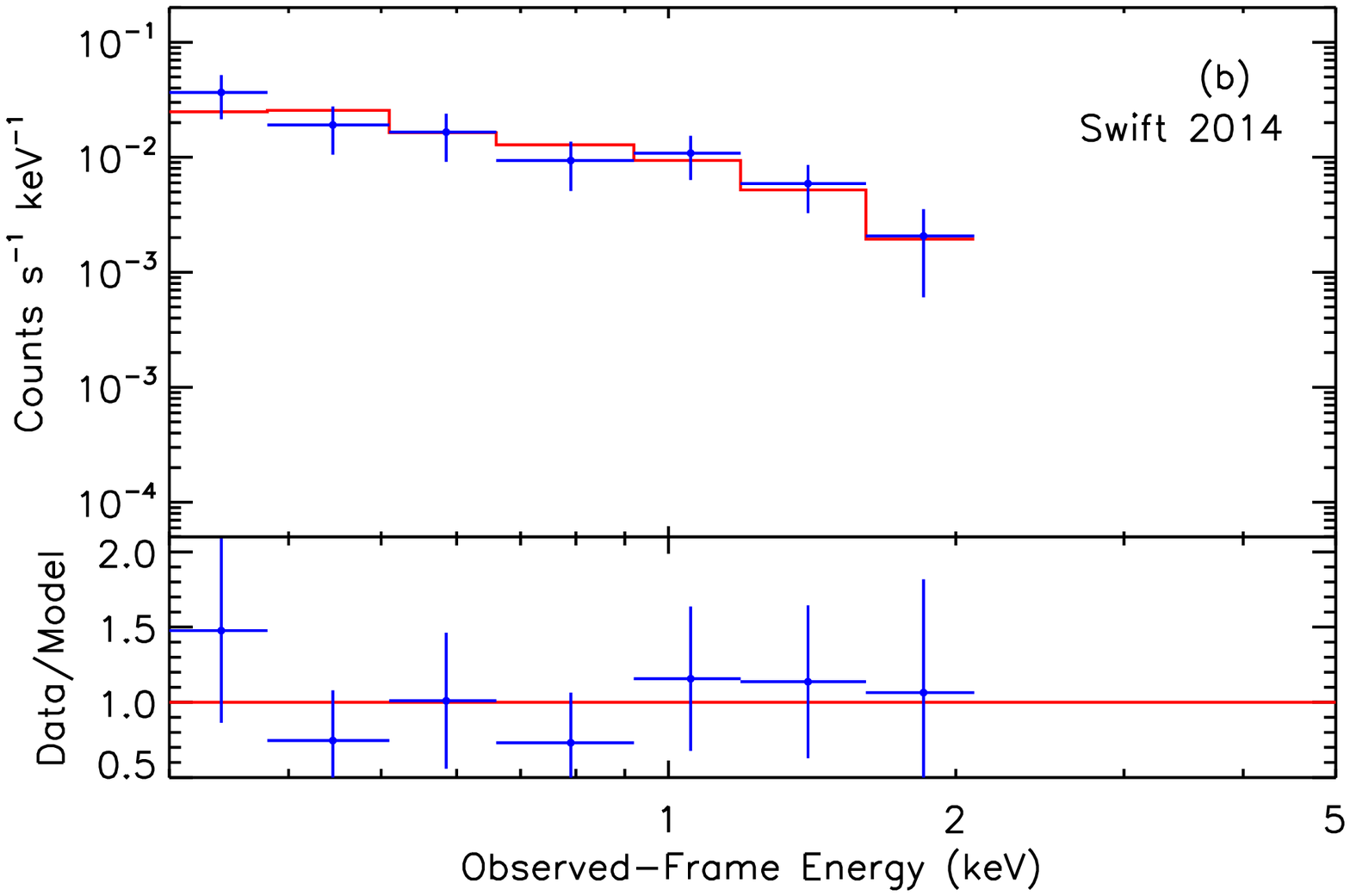}
  \includegraphics[scale=0.42]{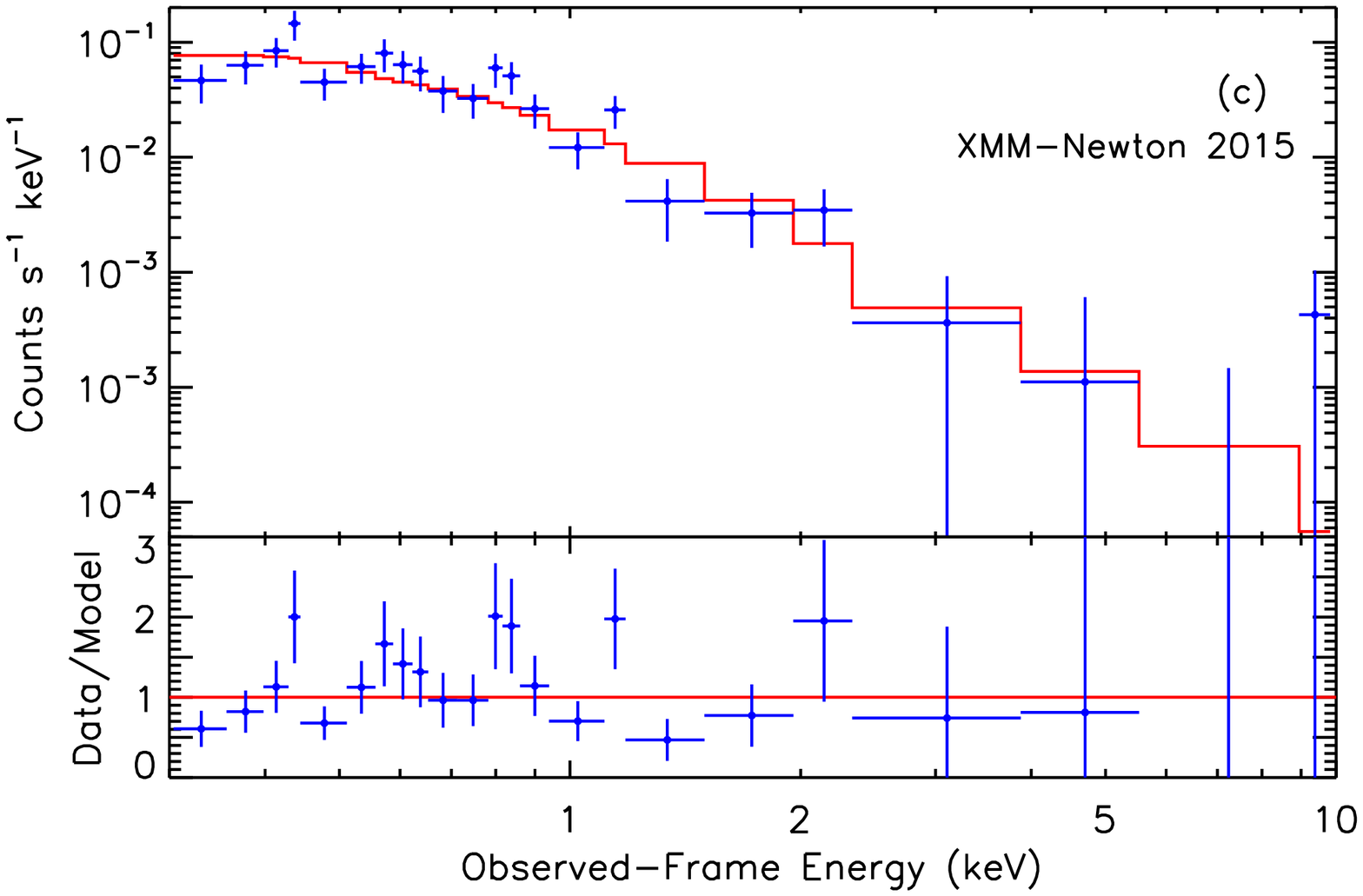}
\caption{\xray\ spectra overlaid with best-fit models for the (a)
 2013 \swift, (b) 2014 \swift, and (c) 2015 \xmm\ observations.
The three \hbox{0.3--10~keV} spectra were fitted with a single
 power-law model modified by Galactic absorption.
For display purposes, the spectra are grouped such that each bin
 has at least $\rm 3\sigma$ significance for the 2013 \swift\ and
 2015 \xmm\ spectra, and $\rm 2.2\sigma$ significance for the 2014
 \swift\ spectrum. In each subfigure, the best-fit model is shown
  in red, and the bottom panel shows the ratio of the spectral
 data to the best-fit model. These simple absorbed power-law
  models provide reasonable fits to the \xray\ data, with steep photon
   indices of $\rm \Gamma_s \approx 3$.
\label{fig-spec}}
\end{figure}

\begin{figure}
\centerline{
\includegraphics[scale=0.4]{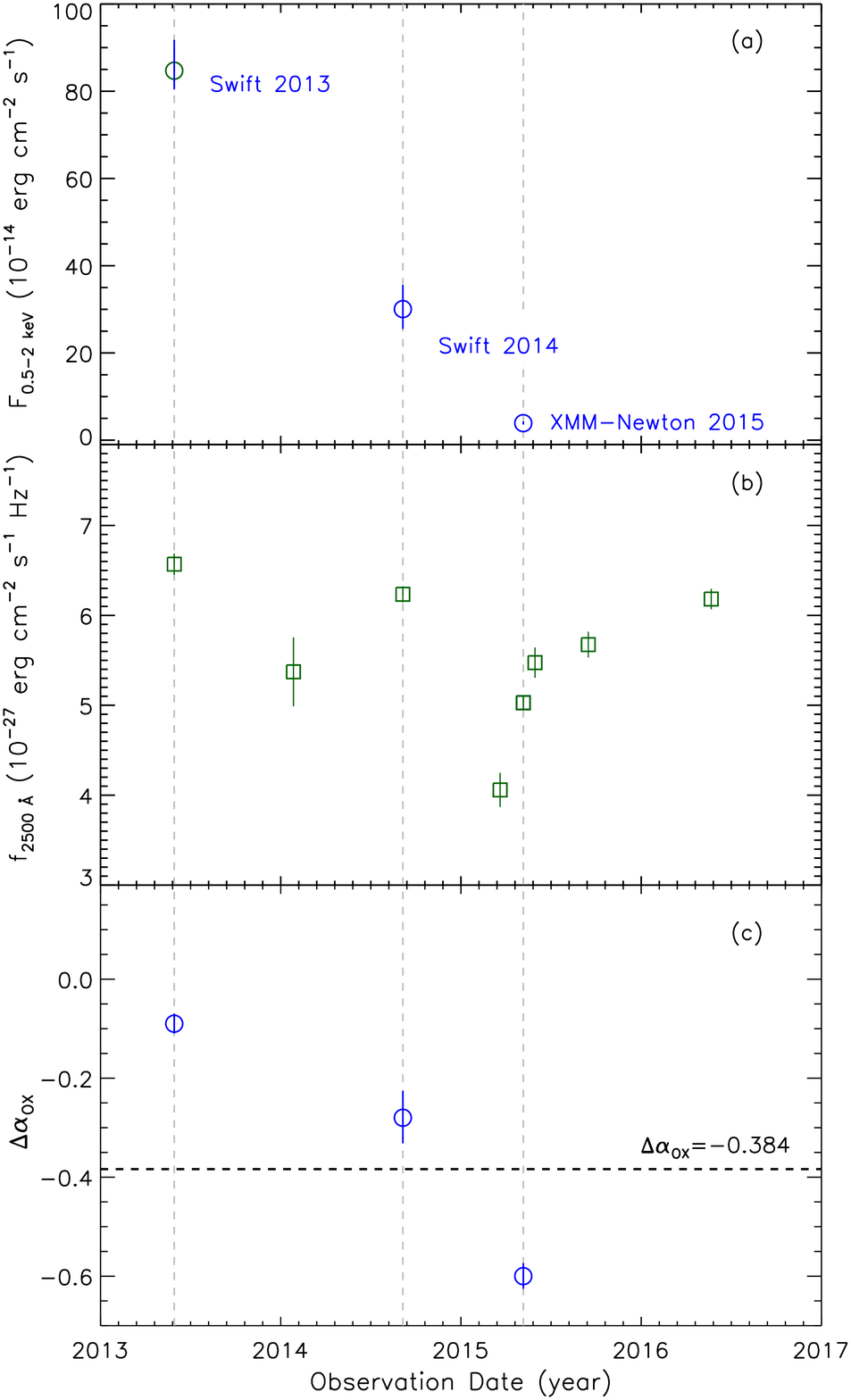}
}
\caption{Light curves of the (a) 0.5--2~keV flux, (b) rest-frame $2500~{\textup{\AA}}$ flux density, and 
(c) $\Delta\alpha_{\rm OX}$ for SDSS $\rm J0751+2914$.
The error bars show the $\rm 1\sigma$ statistical uncertainties. The vertical gray dashed lines mark the dates of the simultaneous \xray\ and  
UV observations.
 In panel (c), the dashed black line ($\Delta\alpha_{\rm OX}=-0.384$) shows a $\approx 2.3\sigma$ deviation from $\alpha_{\rm OX,exp}$, corresponding to an \xray\ weakness factor of $f_{\rm weak}=10$.
}
\label{fig-lc}
\end{figure}

\begin{figure}
\centerline{
\includegraphics[scale=0.42]{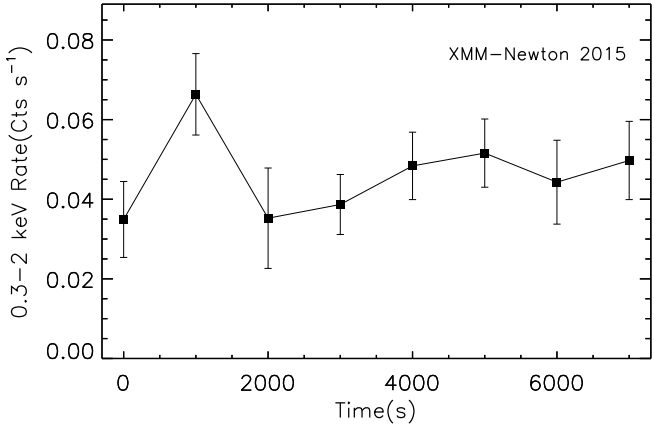}
}
\caption{
The 0.3--2~keV background-subtracted light curve for the 2015 \xmm\ observation. The bin size is 1 ks, corresponding to $\sim 892$~s in the rest frame. 
}
\label{fig-lc_sx}
\end{figure}

\begin{figure}
\centerline{
\includegraphics[scale=0.42]{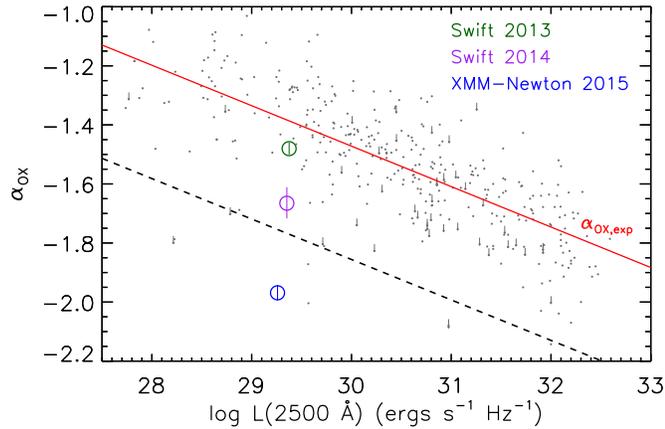}
}
\caption{X-ray-to-optical power-law slope ($\alpha_{\rm OX}$)
vs.~$2500~{\textup{\AA}}$ monochromatic luminosity.
The three open circles represent the measurements from the three \xray\ observations for SDSS~$\rm J0751+2914$.
The small black dots and downward arrows (upper limits) show the typical AGNs in the sample of \cite{Steffen2006},
and the solid red line shows the best-fit \hbox{$\alpha_{\rm OX}$--$L_{\rm 2500~{\textup{\AA}}}$} relation. The dashed black line ($\Delta\alpha_{\rm OX}=-0.384$) shows a $\approx 2.3\sigma$ deviation from the expected \hbox{$\alpha_{\rm OX}$--$L_{\rm 2500~{\textup{\AA}}}$} relation, corresponding to an \xray\ weakness factor of $f_{\rm weak}=10$.
}
\label{fig-aox}
\end{figure}

\begin{figure}
\centerline{
\includegraphics[scale=0.42]{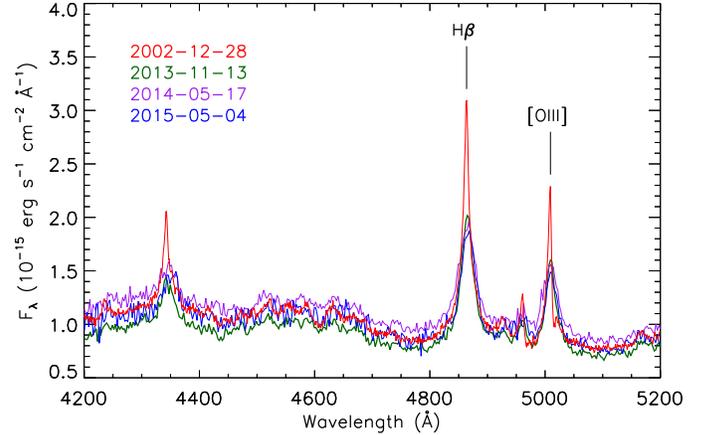}
}
\caption{
SDSS and Lijiang spectra for SDSS~$\rm J0751+2914$. The 2013 and 2014 Lijiang spectra (green and purple) are quasi-simultaneous with the two \swift\ observations, and the 2015 Lijiang spectrum (blue) is simultaneous with the \xmm\ observation.
The emission lines in the SDSS spectrum appear sharper due to the 
higher spectral resolution of the SDSS observation.
}
\label{fig-opt}
\end{figure}

\begin{figure*}
\centerline{
\includegraphics[scale=0.45]{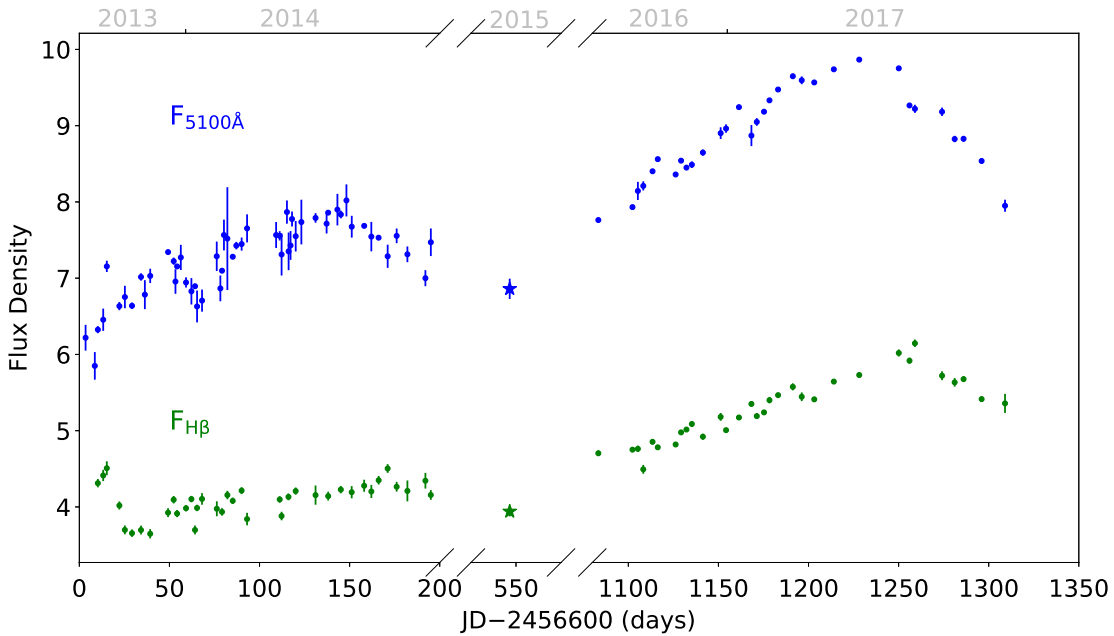}
}
\caption{
Light curves of the $5100~{\textup{\AA}}$ continuum flux density and $\rm H\beta$ emission-line flux (in units of $\rm 10^{-16}~erg~cm^{-2}~s^{-1}~\textup{\AA}^{-1}$ and $\rm 10^{-14}~erg~cm^{-2}~s^{-1}$, respectively) from the Lijiang monitoring observations.
The data points were collected from 
\cite{Du2015,Du2018}, and the stars were measured from the 2015 May 4 Lijiang spectrum, 
which was observed simultaneously with the 2015 \xmm\ observation.
 The 2013 \swift\ observation was observed 170 days before
the first Lijiang observation in 2013 (the first blue data point of
the first segment), and the 2014 \swift\ 
observation was observed 110 days after the last Lijiang observation
 in 2014 (the last blue data point of the first segment).
}
\label{fig-lc_opt}
\end{figure*}

\begin{deluxetable*}{ccccccccccc}
\tablewidth{0pt}
\tablecaption{X-ray and UV/optical properties
}
\tablehead{
\colhead{Observatory}   &
\colhead{Date}   &
\colhead{$f_{\rm 2~keV}$} &
\colhead{$f_{\rm W2}$} &
\colhead{$f_{\rm M2}$} &
\colhead{$f_{\rm W1}$} &
\colhead{$f_{\rm U}$} &
\colhead{$f_{\rm 2500\textup{\AA}}$} &
\colhead{$\alpha_{\rm OX}$} &
\colhead{$\Delta\alpha_{\rm OX}$} &
\colhead{$f_{\rm weak}$} \\
\colhead{(1)}  &
\colhead{(2)}  &
\colhead{(3)}  &
\colhead{(4)}  &
\colhead{(5)}  &
\colhead{(6)}  &
\colhead{(7)}  &
\colhead{(8)}    &
\colhead{(9)}   &
\colhead{(10)}  &
\colhead{(11)}  
}
\startdata
\swift & 2013--05--27&$ 9.13$&--& -- &$6.30$&-- &$ 6.57$&$-1.48\pm0.02$&$ -0.09$& $1.8^{+0.2}_{-0.2}$\\
&2014--01--25&--&$4.74$&-- & -- & --&$5.37$& -- & -- &  -- \\
&2014--09--04&$ 2.86$& --& -- & --&$7.04$&$ 6.23$&$-1.67\pm0.05$&$ -0.28$&  $5.4^{+2.0}_{-1.5}$\\
&2015--03--18&--& --& --&$ 3.89$& --&$ 4.06$&  -- & --& --  \\
&2015--05--28&--&$4.83$&--& -- & --&$5.47$& --  & --  & --  \\
&2015--09--14& -- &--& --&$  5.44$& --&$ 5.67$& --& -- &  -- \\
&2016--05--20&  --  &--&--& --  &$6.98$&$ 6.18$& -- &  --& --  \\
\xmm &  2015--05--04&$ 0.36$& --  &$4.60$&$5.12$& $5.80$ &$ 5.03$&$-1.97\pm0.03$&$ -0.60$&$36.2^{+6.0}_{-5.3}$
\enddata
\tablecomments{
Cols. (1)--(2): \xray\ observatory and observation date.
Col. (3): Galactic absorption corrected flux density at rest-frame 2~keV in units of $\rm 10^{-31}erg~cm^{-2}~s^{-1}~Hz^{-1}$. We used only two of the seven \swift\
 XRT observations.
Cols. (4)--(7): Galactic extinction corrected flux densities in the UVOT and OM bands, given in units of
$\rm 10^{-27}erg~cm^{-2}~s^{-1}~Hz^{-1}$. 
Col. (8): flux density at rest-frame 2500~\AA, derived from the 
UVOT/OM photometric data. 
Col. (9): \xray-to-optical power-law slope parameter. 
The $1\sigma$ uncertainty was propagated from the uncertainty of
$f_{\rm 2~keV}$.
Col. (10): difference between the observed $\alpha_{\rm OX}$ and expected $\alpha_{\rm OX,exp}$ derived from the $\alpha_{\rm OX}\textrm{--}L_{\rm 2500\textup{\AA}}$ relation of \cite{Steffen2006}.
Col. (11): factor of \xray\ weakness in accordance with $\Delta\alpha_{\rm OX}$.
}
\label{tbl-xbasic}
\end{deluxetable*}

\begin{deluxetable*}{cccccccccc}
\tabletypesize{\scriptsize}
\tablecaption{Spectral Fitting Results
}
\tablehead{
\colhead{Observatory}&
\colhead{Observation}&
\colhead{Band} &
\colhead{Total} &
\colhead{Background} &
\colhead{$\rm\Gamma_s$} &
\colhead{$\rm\Gamma_h$} &
\colhead{C-stat/dof} &
\colhead{$F_{\rm 0.5\textrm{--}2~keV}$} &
\colhead{$F_{\rm 2\textrm{--}10~keV}$} \\
\colhead{   } &
\colhead{ Date }&
\colhead{(keV)}  &
\colhead{Counts } &
\colhead{Counts } &
\colhead{    }  &
\colhead{     }  &
\colhead{   } &
\colhead{($\rm 10^{-13} erg~cm^{-2}~s^{-1}$) }  &
\colhead{($\rm 10^{-13} erg~cm^{-2}~s^{-1}$)}
}
\startdata
\swift\ &2013--05--27& 0.3--10& 176& 3.4&$     2.89_{-0.13}^{+0.14}$& -- & $108.7/103$&$     8.47_{-0.43}^{+0.71}$&$3.03_{-0.46}^{+0.62}$\\
        &            & 1.8--10& 22&1.8 &--& $     2.36_{-0.69}^{+0.72}$&$22.9/22$&$     6.70_{-5.67}^{+2.87}$&$5.15_{-0.66}^{+2.70}$\\
       \swift\ &2014--09--04   & 0.3--10&34& 2.0&$     3.07_{-0.37}^{+0.37}$& -- &$20.5/30$ &$     3.01_{-0.45}^{+0.55}$&$0.83_{-0.33}^{+0.64}$\\
\xmm\   &2015--05--04 & 0.3--10&295&72.1 &$     3.06_{-0.17}^{+0.18}$& --  &$126.7/178$&$     0.39_{-0.03}^{+0.04}$&$     0.11_{-0.03}^{+0.04}$
 \enddata
\tablecomments{The spectral fitting model adopted is a single 
power-law model modified by Galactic absorption ({\sc wabs*zpowerlw}). 
The Galactic absorption ({\sc wabs}) is fixed at 
$N_{\rm H}=3.55\times 10^{20}~\rm cm^{-2}$ \citep{Kalberla2005}.
All quoted errors are at a 68\% ($1\sigma$) confidence level.
}
\label{tbl-fit}
\end{deluxetable*}

\subsection{X-ray Spectral Analysis}\label{subsec:xspec}
All of the \xray\ spectral fitting was performed with XSPEC 
\citep[v12.9.1;][]{Arnaud1996}. Due to the small numbers of counts 
in the spectra, the Cash statistic (CSTAT; \citealt{Cash1979})
\footnote{The W statistic was actually used in the XSPEC spectral 
fitting when background spectra are included. 
see https://heasarc.gsfc.nasa.gov/xanadu/xspec/manual/
XSappendixStatistics.html for details.} 
was used in parameter estimation as it is based on the Poisson
 distribution. We first used a simple power-law model modified by 
 Galactic absorption ({\sc wabs*zpowerlw}) to fit the 0.3--10~keV 
 spectra. The Galactic neutral hydrogen column density was fixed at
 $N_{\rm H}=3.55\times10^{20}\rm~cm^{-2}$ \citep{Kalberla2005}.
The \hbox{best-fit} results are presented in Figure~\ref{fig-spec} and
 the model parameters are reported in Table~\ref{tbl-fit}.
 The error bars plotted in Figure~\ref{fig-spec} and the parameter errors listed in Table~\ref{tbl-fit}   
are at a 68\% ($1\sigma$) confidence level.
In general, these models provide reasonable fits to the three spectra.
The three power-law photon indices ($\rm \Gamma_s$ in Table~\ref
{tbl-fit}) indicate steep spectral shapes, which are consistent with 
those of NLS1s \citep[e.g,][]{Brandt1997,Leighly1999b}.

Soft \xray\ excess emission can steepen the power-law slope of an
\xray\ spectrum. It is common in NLS1s, and also in quasars 
in general \citep[e.g.,][]
{Boller1996,Leighly1999b,Porquet2004,Grupe2010,Marlar2018}. 
Thus, we also attempted to fit the 1.8--10~keV (rest-frame energies
 above 2~keV) hard \xray\ spectra with the same power-law model to 
 isolate the intrinsic \xray\ emission from the corona.
This analysis is only feasible for the 2013 \swift\ 
 observation, since there is no photon in the hard \xray\ spectrum 
 of the 2014 \swift\ observation, and the hard \xray\ spectrum 
 of the 2015 \xmm\ observation is dominated by the $\approx 42$ 
 background photons (leaving $\approx 8$ net source photons).
Due to the limited \hbox{signal-to-noise} ratio of the 2013 \swift\
 hard \xray\ spectrum, its power-law photon index 
($\rm \Gamma_h$ in Table~\ref{tbl-fit}) derived from the fitting 
has considerable uncertainties.
Considering the large uncertainties,
the $\rm \Gamma_h$ value is consistent with the $\rm \Gamma_s$ value
of the entire $0.3\textrm{--}10$~keV spectrum, 
and the rest-frame 2~keV flux density derived 
from the $1.8\textrm{--}10$~keV spectral fitting is also consistent
 with that derived from the $0.3\textrm{--}10$ keV spectral fitting.
 Since the spectra appear to be dominated by soft-band counts, we also fit the same single 
 power-law model to the $0.3\textrm{--}1.8$~keV 
 spectra of the 2013 \swift\ and 2015 \xmm\ observations. 
 The fitting revealed two steep power laws of $\Gamma\approx 3$, consistent with
 the $0.3\textrm{--}10$~keV spectral fitting results.
In the discussion below, we thus adopt the 2~keV flux densities and 
$0.5\textrm{--}2$~keV fluxes derived from the 
$0.3\textrm{--}10$~keV spectral fitting.

\subsection{X-ray Variability}\label{subsec:xvar}
Figure~\ref{fig-lc}(a) depicts the long-term variability of the 
\hbox{0.5--2~keV} flux for \hbox{SDSS~$\rm J0751+2914$}.
With the flux dropping by a factor of $21.6\pm 2.4$ (factors of 
$2.8\pm 0.5$ and $7.7\pm 1.5$ successively from 2013 to 2014 
and 2014 to 2015), \hbox{SDSS~$\rm J0751+2914$} has gradually 
fallen into a low \xray\ flux state.
In addition, we examined the short-term \xray\ variability of 
\hbox{SDSS~$\rm J0751+2914$} within the 2013 \hbox{normal-flux} state
and 2015 low-flux state.
Figure~\ref{fig-lc_sx} shows the $0.3\textrm{--}2$~keV light curve 
with time bins of 1~ks for the 2015 \xmm\ observation. 
The average count rate in this observation is 0.046~$\rm Cts~s^{-1}$, 
with a \hbox{root-mean-square} (rms) variability of 0.011~$\rm
 Cts~s^{-1}$ and a fractional rms variability amplitude 
 \cite[e.g., Equation 10 of][]{Vaughan2003} of $8\%$.
For the 2013 \swift\ observation,
we extracted a $0.3\textrm{--}2$~keV light curve with time bins  
of 500~s.
The measured average count rate is 0.055~$\rm Cts~s^{-1}$, with 
a rms variability of 0.020~$\rm Cts~s^{-1}$ and 
a fractional rms variability amplitude of $32\%$. 

With the simultaneous \xray\ and UV observations, we can 
compute reliably the X-ray-to-optical power-law slope
($\alpha_{\rm OX}$) of SDSS~$\rm J0751+2914$. 
We first
measured a UV/optical spectral slope using the
data of the 2015 \xmm\ OM observation, which is the only
observation having multiple filters 
(see Table~\ref{tbl-obs}).
Fitting a single power-law model to the U, UVW1, and UVM2 data points revealed a spectral 
slope of $\alpha_\nu= -0.57$. This slope is 
consistent with those of typical NLS1s
\citep[e.g.,][]{Grupe2010}. We then determined the 
$2500~{\textup{\AA}}$ flux density of the 2015 \xmm\ 
observation from the best-fit model of the OM data.
For each of the seven \swift\ UVOT observations, only one UV 
filter was used.
The $2500~{\textup{\AA}}$ flux densities were extrapolated from the flux densities of the available filters, adopting the same power-law slope of $\alpha_\nu= -0.57$.  
If we adopt a spectral slope of $\alpha_\nu= -0.44$ in the extrapolation which is the average value for typical quasars \citep[e.g.,][]{Vanden2001}, the resulting $\alpha_{\rm OX}$ values would only change slightly (by less than 0.01) and our following analyses and discussions would not be affected. After obtaining the $\alpha_{\rm OX}$ values (see Table 2), we calculated the difference ($\Delta\alpha_{\rm OX}$) between the observed $\alpha_{\rm OX}$ and that ($\alpha_{\rm OX,exp}$) expected from the $\alpha_{\rm OX}$--$L_{2500~{\textup{\AA}}}$ relation, which indicates the level of \xray\ weakness. 
These \xray\ and UV/optical properties of SDSS~$\rm J0751+2914$ are listed in Table~\ref{tbl-xbasic}.

Figure~\ref{fig-aox} shows the $\alpha_{\rm OX}$ versus 
$L_{2500~{\textup{\AA}}}$ values of the three \xray\ 
observations for SDSS $\rm J0751+2914$.
Typical AGNs in the sample of \cite{Steffen2006} are also 
presented for comparison.
The 2013 \swift\ data point of \hbox{SDSS~$\rm J0751+2914$} is
 close to the red line that represents the 
 $\alpha_{\rm OX}$--$L_{2500~{\textup{\AA}}}$ relation in
  \cite{Steffen2006}, indicating a normal \xray\ emission 
  level at this time. However, the quasar became extremely 
  \xray\ weak in the 2015 \xmm\ observation with 
$\Delta\alpha_{\rm OX}=-0.60\pm0.03$ ($f_{\rm weak}=36.2^{+6.0}_{-5.3}$), 
 corresponding to a $3.6\sigma$ deviation from the 
  $\alpha_{\rm OX}$--$L_{2500~{\textup{\AA}}}$ relation 
  (see Table~5 of \citealt{Steffen2006}). 
The $\Delta\alpha_{\rm OX}$ variability is also shown in 
Figure~\ref{fig-lc}(c); the decreasing of 
the $\Delta\alpha_{\rm OX}$ value from 2013 to 2015 is coordinated 
with the drop of the \xray\ flux.

\subsection{UV Variability}\label{subsec:uvvar}
Presented in Figure~\ref{fig-lc}(b) is the light curve of the
 $2500~{\textup{\AA}}$ flux density derived from the UVOT and OM 
 photometric data. The gray dashed lines indicate the dates of the 
 simultaneous \xray\ and UV observations.
Combined with the results reported in Table~\ref{tbl-xbasic},
we find that the UV flux density has a much smaller variability 
amplitude compared to the 2~keV flux density.
The $2500~{\textup{\AA}}$ flux density varied by just a factor of 
1.3 between the high \xray\ flux state and the low \xray\ flux state.
Such little variation of the UV flux suggests that the physical 
mechanism leading to the strong \xray\ variability largely does 
not affect the UV emission.

We have no UV spectrum of SDSS~$\rm J0751+2914$ to 
 identify whether it is a BAL quasar, in which case it may be
 affected by absorption associated with outflows 
 \citep[e.g.,][]{Murray1995,Matthews2016}.
 BAL quasars with extreme \xray\ variability
 often show significant 
 UV/optical continuum and BAL variability coordinated with 
 the \xray\ variability \citep[e.g.,][]{Gallagher2004,Saez2012,Kaastra2014,Mehdipour2017}.
SDSS~$\rm J0751+2914$ lacks significant UV/optical 
 variability coordinated with its \xray\ variability. It is thus 
 probably not a BAL quasar with strong \xray\ variability.  
A UV spectroscopic observation of SDSS~$\rm J0751+2914$ is required 
to confirm this notion.

\subsection{Optical Spectrum and Light Curves}\label{subsec:opt}
Three Lijiang spectra in the rest-frame $4200\textrm{--}
5200~{\textup{\AA}}$ range are shown in Figure~\ref{fig-opt}, which
 also includes the SDSS spectrum observed on 2002 December 28.
The continuum and emission lines did not vary significantly in general
 with a $\sim 10\%$ percent variability amplitude among the four 
 observations.
The emission-line profiles in the SDSS spectrum appear relatively
 sharp because of the better spectral resolution of the SDSS 
 observation.

Figure~\ref{fig-lc_opt} shows the light curves of the
$5100~{\textup{\AA}}$ continuum flux density 
($F_{5100~{\textup{\AA}}}$) and the $\rm H\beta$ emission-line flux
 ($F_{\rm H\beta}$) during the \hbox{2013--2014} and \hbox{2016--2017} 
 RM monitoring periods \citep{Du2015,Du2018}.
We also added a data point measured from the 2015 May 4 Lijiang
 spectrum following the approach in \cite{Du2014}.
 During 2013--2015,
 the $F_{5100~{\textup{\AA}}}$ and 
$F_{\rm H\beta}$ values of \hbox{SDSS $\rm
 J0751+2914$} have maximum variability amplitudes of $\sim 
 30\%$, which are only mild compared to its extreme \xray\
  variability. This indicates that the accretion 
  rate of \hbox{SDSS $\rm J0751+2914$} did not change 
  significantly during this period and the extreme \xray\ variability should be driven by some other mechanisms.
 The $F_{5100~{\textup{\AA}}}$ and 
$F_{\rm H\beta}$ values in the \hbox{2016--2017} period have 
increased in
 general compared to those in the 2013--2015 period, 
   and the maximum variability amplitudes of the two parameters 
   are $\sim70\%$ among all the RM observations.
 This variability is more significant than that of the other SEAMBHs 
 which generally have maximum variability amplitudes of 
 $\sim20\textrm{--}30\%$ \citep[see][]{Du2014,Du2015,Du2016,Du2018}.
 The \xray\ observations of 
 \hbox{SDSS $\rm J0751+2914$} were performed in 2013--2015,
 and there is no coordinated optical continuum
and emission-line variability with the \xray\ variability of
\hbox{SDSS $\rm J0751+2914$} during this period.

 We also investigated the $V$-band light curve of 
 \hbox{SDSS $\rm J0751+2914$} obtained from the Catalina 
 \hbox{Real-Time}
 Transient Survey (CRTS; \citealt{Drake2009}). 
 The monitoring period is between 2005 April and 2013 September.
 During this period, the $V$-band magnitude of 
 \hbox{SDSS $\rm J0751+2914$} varied between 
 $15.89\textrm{--}16.27$ (a 42\% maximum variability amplitude in
  flux), with a mean value of 16.06.

\subsection{Mutiwavelength Spectral Energy Distribution}
\begin{figure}
\centerline{
\includegraphics[scale=0.42]{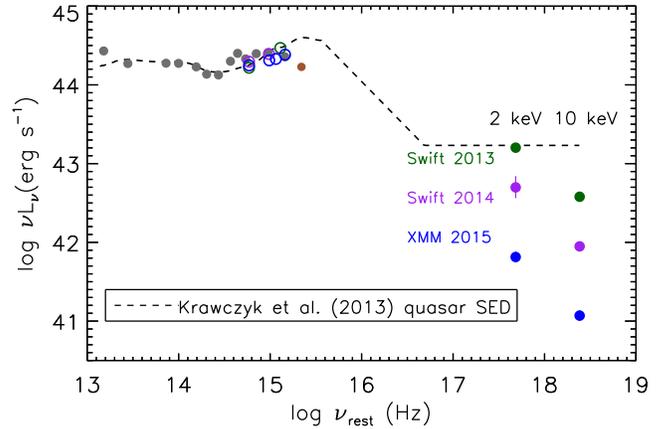}
}
\caption{Rest-frame IR-to-X-ray SED for SDSS $\rm J0751+2914$.
The \hbox{IR-to-UV} photometric data points were gathered from
the {\it WISE}, 2MASS, SDSS, and {\it GALEX} catalogs.
The UVOT and OM photometric data and the $5100~{\textup{\AA}}$
monochromatic luminosities are shown as open circles.
The 2~keV and 10~keV monochromatic luminosities from the two
\swift\ and one \xmm\ observations are shown as points with
different colors (green, purple, and blue).
Data points from simultaneous or quasi-simultaneous \xray\
and UV/optical observations are represented with the same
color (green, purple, or blue). 
The dashed line shows the mean quasar SED of
low-luminosity SDSS quasars \citep{Krawczyk2013}, which is
scaled to the mean $5100~{\textup{\AA}}$ luminosity of SDSS
$\rm J0751+2914$. 
}
\label{fig-sed}
\end{figure}

 We gathered infrared (IR)-to-UV photometric data to
construct the rest-frame spectral energy distribution (SED)
 for SDSS $\rm J0751+2914$, which is shown in
 Figure~\ref{fig-sed}.
The data were collected from the public catalogs of
the {\it Wide-field~Infrared~Survey~Explorer} \cite[{\it
WISE};][]{Wright2010}, Two Micron All Sky Survey \cite[2MASS;]
[]{Skrutskie2006}, SDSS, and {\it Galaxy~Evolution~Explorer}
 \cite[{\it GALEX};][]{Martin2005}.
We added the UVOT and OM photometric data, and the
 corresponding 2~keV and 10~keV luminosities to the SED. 
 Also, the $5100~{\textup{\AA}}$ monochromatic
 luminosities calculated from the three Lijiang spectra in
 Figure~\ref{fig-opt} were added.
 We caution that most of these photometric data are not 
 contemporaneous. All the SED data were corrected for the Galactic 
 extinction at the source position.
The mean SED of typical SDSS quasars with luminosities of 
log$(\nu L_\nu|_{\lambda=2500~\textup{\AA}} / \rm erg~s^{-1}) \le 45.41$ 
in \cite{Krawczyk2013}, scaled to the mean $5100~{\textup{\AA}}$
luminosity of SDSS~$\rm J0751+2914$,
is shown in Figure~\ref{fig-sed} for comparison.
We note that the IR-to-UV SED of SDSS~$\rm J0751+2914$ is 
 consistent with those of typical quasars, except for the {\it GALEX} 
 FUV data (shown as the brown point in Figure~\ref{fig-sed})
  at \hbox{rest-frame} $1364.2~{\textup{\AA}}$ ($\rm log~\nu_{rest}
 =15.34$), which lies below the mean SED.
The date of this {\it GALEX} observation is 2006 December 29,
which is prior to the three \xray\ observations and the RM 
observations, and thus we cannot determine whether this particular
feature is related to the extreme \xray\ variability.
A far UV spectrum is required to examine if
there is any UV absorption. The \xray\ data points indicate
the soft \xray\ spectral shapes, which have not changed
significantly between the three \xray\ observations.
In spite of the extreme \xray\ flux variability,
the optical-to-UV SED did not change significantly.

We estimated the bolometric luminosity of SDSS $\rm
J0751+2914$ by integrating the scaled SED template of
\cite{Krawczyk2013} shown in Figure~\ref{fig-sed}.
Most of the IR radiation (\hbox{$\sim1 \textrm{--} 30~\rm\mu m$})
is produced in the large-scale "dust torus" beyond the
accretion disk. It is considered to be the reprocessed
emission and should not be included in the computation of
the bolometric luminosity \cite[e.g.,][and references
therethin]{Krawczyk2013}.
However, super-Eddington accreting quasars are expected to
produce much stronger extreme UV radiation than typical quasars
\citep[e.g.,][]{Wang2014b,Castell2016},
but this portion of the SED is not observable, and it is not
 represented by the \cite{Krawczyk2013} mean quasar SED.
Thus we included the IR SED in the integration to
compensate somewhat for the uncertain extreme UV emission.
The resulting bolometric luminosity is $\rm 1.66\times10^{45}
~erg~s^{-1}$ ($\rm 1.03\times10^{45}~erg~s^{-1}$ if not including 
the \hbox{$1 \textrm{--} 30~\rm\mu m$} SED). 
We note that the \xray\ spectrum in the SED template 
was included in the integration of the bolometric luminosity.
 Since the $2\textrm{--}10$~keV template luminosity contributes only a small fraction 
($\approx 2\%$) of the bolometric luminosity, the result would
only change slightly if we adopt the $2\textrm{--}10$~keV luminosities determined from the 
observational data. We also caution that the observed \xray\ 
luminosities likely do not represent the intrinsic \xray\
 luminosity (see Section~\ref{sec:model}).

We also estimated the bolometric luminosity using the bolometric
correction from the $3~\rm\mu m$ monochromatic luminosity
 \citep{Gallagher2007}, which was derived from the {\it WISE}
  photometric data. The resulting bolometric luminosity is 
 $\rm 1.61\times10^{45}~erg~s^{-1}$, consistent with the integrated 
 luminosity of the $30~\rm\mu m\textrm{--}10~keV$ SED. 
Given that the BH mass of SDSS
$\rm J0751+2914$ is $1.6\times10^7~M_\odot$,
we obtained an Eddington ratio of 0.7. This high Eddington ratio 
suggests that SDSS~$\rm J0751+2914$ is indeed accreting at a high
 accretion rate.   
We caution that the BH mass and subsequently the Eddington ratio 
may have substantial uncertainties (see Section~\ref{sec:intro}).

\section{DISCUSSION}\label{sec:discuss}
\subsection{Extremely X-ray Variable AGNs}\label{subsec:sample_ex}
SDSS $\rm J0751+2914$ is another NL type 1 quasar with extreme \xray\ 
variability. 
 It varied in X-rays by a factor of $21.6$ on a
 timescale of three years. In the same period 
 there was no coordinated UV/optical continuum or emission-line 
 variability, indicating that the accretion rate of SDSS~$\rm J0751+2914$ is almost constant and there are some other factors instead of a change of accretion rate driving the extreme \xray\ variability.
  These features are similar to the other three 
  extremely \xray\ variable NL type 1 quasars reported, 
  PG~$0844+349$ \citep[e.g.,][]{Gallo2011}, 
  PG~$1211+143$ \citep[e.g.,][]{Bachev2009}, 
  and PHL~1092 \citep[e.g.,][]{Miniutti2012}.
  Thus SDSS $\rm J0751+2914$ is a new member of 
  the extremely \xray\ variable quasar population.
The basic characteristics of these four  
extremely \xray\ variable quasars are listed in 
Table~\ref{tbl-sample}.
 
Table~\ref{tbl-sample} also lists a few representative
 extremely \xray\ variable NLS1s (with $L_{\rm 5100~\textup{\AA}}<
  10^{44}~\rm erg~s^{-1}$) selected from the literature.  
We note that the \hbox{BH-mass} estimates and the computed 
bolometric luminosities have large uncertainties.   
 Thus there are substantial uncertainties on the derived 
 $\rm\dot{\mathscr{M}}$ and $\lambda_{\rm Edd}$ values 
 (see Section~\ref{sec:intro}).
 In general, these extremely \xray\ variable AGNs have high 
 accretion rates. According to the criterion of 
 $\rm\dot{\mathscr{M}}>3$ used for identifying SEAMBH candidates
 \citep{Du2014,Du2015}, all of these extremely \xray\ variable 
 AGNs (except IRAS~$13224-3809$) can be considered as SEAMBH 
 candidates. This result implies a connection between 
 extreme \xray\ variability and high accretion rates in AGNs. 

The steep \xray\ spectral shapes of SDSS $\rm J0751+2914$, 
 whether in the high or low \xray\ flux state, are similar to
  those of PHL~1092 \citep{Miniutti2009,Miniutti2012}. 
  Considering the large uncertainties, the spectral shape 
  ($\rm \Gamma_s$ in Table~\ref{tbl-fit}) of SDSS~$\rm J0751+2914$ 
  did not change significantly. 
However, a flattening of the hard (\hbox{$> 2~\rm keV$}) \xray\ 
spectral shape from the normal to low state is often observed in 
other extremely \xray\ variable AGNs (PG~$0844+349$: 
\citealt{Gallo2011}; PG~$1211+143$: \citealt{Bachev2009}; 
Mrk~335: \citealt{Grupe2012,Gallo2015}; 1H~$0707-495$: 
 \citealt{Fabian2012}; IRAS~$13224-3809$: \citealt{Jiang2018}).
 Their low-state \xray\ spectra show substantial curvature in the
$\approx~2\textrm{--}6$~keV band,
which is generally interpreted as blurred reflection arising 
within a few gravitational radii of the BH or partial covering
 absorption (see discussion in Section~\ref{sec:model} below).
 For SDSS~$\rm J0751+2914$ and PHL~1092 that are at relatively 
 high redshifts and have relatively low \xray\ fluxes, 
 the current observations are probably not sufficiently 
 sensitive to detect this hard \xray\ component. 
 Their low-state spectra are likely dominated by the soft 
 \xray\ excess component (see Figure~6 of \citealt{Miniutti2012} 
 and Figure~\ref{fig-spec}), and there is no apparent curvature 
 nor hardening emerging in their $\approx 2\textrm{--}10$~keV
 spectra. Hard \xray\ observations with \nustar\ 
 \citep{Harrison2010} and {\it Suzaku} \citep{Mitsuda2007} of 
 some extremely \xray\ variable AGNs suggest that their
 \hbox{$>10$~keV} X-ray fluxes and spectral shapes are less
 variable \citep[e.g.,][]{Gallo2015,Kara2015,Jiang2018}. 
 An additional spectral curvature at higher energies ($\approx 
 20\textrm{--}30$~keV) is also usually observed in different flux
  states, and it is interpreted as the Compton-reflection hump. 
 A hard \xray\ observation with \nustar\ or a deep \xmm\ 
  or \chandra\ observation on SDSS $\rm J0751+2914$
 in its low \xray\ flux state is required to investigate whether
 it possesses a hard/flat \xray\ spectrum similar to 
 the other extremely \xray\ variable AGNs.

\subsection{Occurrence Rate of Extreme X-ray Variability among 
AGNs with High Accretion Rates 
}\label{subsec:frac}

Inspired by the possible connection between extreme \xray\
 variability and high accretion rates,  
we investigated the occurrence of extremely
variable \xray\ sources among AGNs with high accretion rates 
($\lambda_{\rm Edd}\ga 0.1$).
We first target NLS1s that generally have high accretion rates.
In the soft \xray\ selected Seyfert sample of \cite{Grupe2010},
there are 49 broad line Seyfert 1 galaxies (BLS1s) and 43 NLS1s 
(19 of the 43 can be considered as quasars, with $5100~\textup{\AA}$
 luminosities exceeding $\rm 10^{44}~erg~s^{-1}$).
All the NLS1s in this sample are high accretion rate AGNs with 
$\lambda_{\rm Edd}\ga 0.1$ \citep{Grupe2010}.
Only three NLS1s (\hbox{RX~J$2217.9-5941$}:
 \citealt{Grupe2004a}; Mrk~335: \citealt{Grupe2007a}; PG~$1211+034$:
\citealt{Bachev2009}) have been found to vary in \hbox{X-rays} by
factors of more than 10 between multiple \swift\ observations.
There is no BLS1 in this sample found to show 
extreme \xray\ variability.

The fraction of extremely \xray\ variable AGNs ($P_{\rm var}$)
among NLS1s should be larger than $3/43$ ($7\%$), as some of them were 
probably not identified due to the limited number of observations
available. This fraction $P_{\rm var}$ also represents the probability of a NLS1 being extremely
 \xray\ variable. We estimated $P_{\rm var}$ using the 
observations and $\Delta\alpha_{\rm OX}$ values of the NLS1 sample presented in \cite{Grupe2010}. We computed  
the likelihood ($L$) of observing three extremely \xray\
 variable objects among the 43 NLS1s as a function of $P_{\rm var}$, which can be expressed as
\begin{equation}
\begin{aligned}
L\propto P_{\rm var}^{3}
	\times\prod_{i=1}^{40}[P_{\rm var}(1-D)^{n_{{\rm obs},i}}+(1-P_{\rm var})] \ ,
\end{aligned}
\end{equation} 
where $n_{{\rm obs},i}$ is the total number of observations for each
object (with a range of \hbox{1--9} from \citealt{Grupe2010}), and
 {\it D} is the duty cycle of the extremely \xray\ weak state 
 ($\rm \Delta \alpha_{OX} < -0.384$).
 The first term of Equation~(1) corresponds to the likelihood of observing three objects being extremely \xray\ variable.
  The second term corresponds to the 
 likelihood of observing 40 
 objects being \xray\ normal in all $n_{{\rm obs},i}$ 
 observations (either being extremely \xray\ variable but not observed in the \xray\ weak state or being 
 non-variable).
 With the published data, we estimated the duty cycles of the  
 extremely \xray\ weak state of two extremely \xray\ variable 
 AGNs, PHL~1092 and Mrk~335 \citep{Grupe2012,Miniutti2012}. 
 Mrk~335 was in the extremely \xray\ weak state in  
 about $30\%$ of the about four-year long continuous 
 monitoring observations 
 with a total exposure time of $\approx301$~ks 
 \citep[see][]{Grupe2012}.
 PHL~1092 was in the extremely \xray\ weak state 
 in about $60\%$ of the observations \citep[see][]
 {Miniutti2012}\footnote{This fraction is an overestimate
 of the duty cycle of the extreme \xray\ weak state for PHL~1092, 
 as some \xray\ observations of PHL~1092 were follow-up 
 observations triggered by its \xray\ weak state.}.
We thus adopted a $30\textrm{--}60\%$ range for the duty cycle 
of the extremely \xray\ weak state for every extremely \xray\ 
variable AGN. 

\begin{figure}
\centerline{
\includegraphics[scale=0.42]{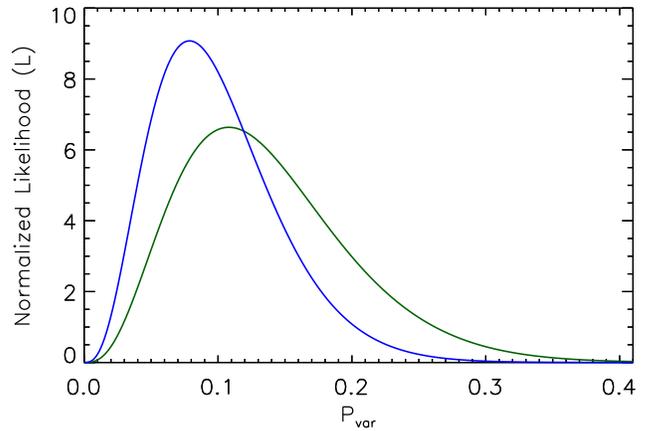}
}
\caption{Likelihood of observing three extremely \xray\ 
variable AGNs in the NLS1 sample of \cite{Grupe2010}
as a function of the occurrence rate of extremely 
\xray\ variable AGNs among NLS1s. The green (blue) curve corresponds to a $30\%$ 
($60\%$) duty cycle of the extremely \xray\ weak state adopted 
in the computation. 
}
\label{fig-p}
\end{figure}

The distributions of {\it L} 
as a function of $P_{\rm var}$ are shown as the green 
(for a duty cycle of 
$30\%$) and blue (for a duty cycle of $60\%$) curves 
in Figure~\ref{fig-p}, which show two peaks at 
$P_{\rm var}\approx 11\%$ and $8\%$, 
respectively, indicating that the most probable $P_{\rm var}$
 value is $11^{+9}_{-3}\%$ ($8^{+7}_{-2}\%$) when the duty 
 cycle of the extremely \xray\ weak state is $30\%$ ($60\%$);
  the $1\sigma$ uncertainties on the $P_{\rm var}$ values were 
 derived from the {\it L} distributions. Thus the fraction of extremely variable \xray\ sources
 among NLS1s is estimated to be $\approx (8\textrm{--}11)\%$.
  The above estimate of $P_{\rm var}$ depends on the uncertain estimate of the duty cycle ({\it D}) of the extremely \xray\ weak state, but the dependence is not very strong. For example, in the extreme cases of $D=10\%$ and $D=100\%$, which are unlikely given the current observations of the limited sample, the corresponding $P_{\rm var}$ values are $24\%$ and $7\%$, respectively. Thus, likely only a small fraction of NLS1s are extremely \xray\ variable.
   
\begin{figure}
\centerline{
\includegraphics[scale=0.42]{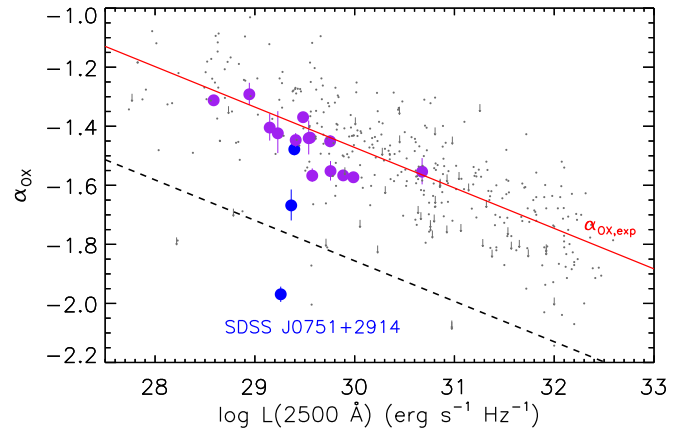}
}
\caption{X-ray-to-optical power-law slope ($\alpha_{\rm OX}$)
vs.~$2500~{\textup{\AA}}$ monochromatic luminosity, similar to 
Figure~\ref{fig-aox}. \hbox{SDSS~$\rm J0751+2914$} is shown 
as the blue dots. The purple dots show the other 15 SEAMBHs 
in \cite{Du2014,Du2015,Du2016,Du2018}, of which we 
analyzed the \xray\ data.
}
\label{fig-quasar}
\end{figure}

Subsequently, we investigated the fraction of extremely variable 
\xray\ sources among the SEAMBHs in the RM campaign conducted
by \cite{Du2014,Du2015,Du2018}, which is the parent sample 
including SDSS~$\rm J0751+2914$.
There are 24 SEAMBHs in this RM campaign, of which 20 have archival 
\xray\ observations. One object, Mrk~486 (PG~1535+547), 
is an \xray\ weak quasar that shows extreme \xray\ spectral 
variability \citep[e.g.,][]{Schartel2005,Ballo2008}. 
 It was classified as a mini-BAL quasar 
 \citep{Brandt2000,Sulentic2006}, so that its extreme \xray\ 
 behavior is likely related to the outflowing wind \citep[e.g.,][]
 {Giustini2016}. We thus excluded it from this SEAMBH sample.
Four objects (Mrk~335, Mrk~142, Mrk~493, and Mrk~1044) of the 
remaining 19 objects are also in the NLS1 sample of \cite{Grupe2010} 
discussed above, and we adopted their \xray\ analysis results. 
Only Mrk~335 among these four objects has been found to show
extreme variability in \hbox{X-rays}. We then derived the \xray\ 
properties of the other 15 AGNs (including 14 quasars and 
one NLS1). The details of the data analysis will be presented in 
H. Liu et al (in preparation). 
Among these 15 AGNs, three objects 
(\hbox{SDSS~$\rm J0751+2914$}, Mrk~382, 
and IRASF~$12397+3333$) have three, five, and three observations,
respectively, and the other 12 objects have been
observed only once.
\hbox{SDSS~$\rm J0751+2914$} is the only object
among these 15 SEAMBHs found to show extreme \xray\ variability.
 The other two objects with multiple observations show little 
 \xray\ variability, and we adopted their mean $\alpha_{\rm OX}$
 values. Figure~\ref{fig-quasar} shows the
$\alpha_{\rm OX}$ versus $L_{2500~{\textup{\AA}}}$
distribution for the 15 objects. 
 Thus two objects (Mrk~335 and \hbox{SDSS~$\rm J0751+2914$}) 
 in the sample of 19 SEAMBHs have ever shown extreme 
 \xray\ variability.
 Based on a process similar to that described above,  
 we estimated that the most probable $P_{\rm var}$ value
is $24^{+24}_{-8}\%$ ($15^{+16}_{-5}\%$) when the duty cycle of the 
extremely \xray\ weak state is $30\%$ ($60\%$).
Thus the fraction of extremely variable \xray\ sources among SEAMBHs 
 is $\approx(15\textrm{--}24)\%$. 
We note that this fraction is not very consistent with the fraction ($8\textrm{--}11\%$) for the NLS1 sample above, probably due to the different selection criteria of the SEAMBH and NLS1 samples which did not yield consistent populations of AGNs with high accretion rate.
In the extreme cases of $D=10\%$ and $D=100\%$, the corresponding $P_{\rm var}$ values for the SEAMBH sample are $58\%$ and $11\%$, respectively.

We investigated if extremely \xray\ variable AGNs also are 
outliers in terms of other physical properties.  
We first compared the dimensionless accretion rates of the two 
extremely variable AGNs discovered (Mrk~335 and \hbox{SDSS~$\rm
 J0751+2914$}) to those of other AGNs in the SEAMBH sample. 
The $\rm log~\dot{\mathscr{M}}$ values of these SEAMBHs span 
a range of \hbox{0.55--2.98} \citep[see][]
{Du2014,Du2015,Du2016,Du2018}. 
Mrk~335 and \hbox{SDSS~$\rm J0751+2914$} both have moderate 
dimensionless accretion rates ($\rm log~\dot{\mathscr{M}}
=1.28~and~1.45$) among all the SEAMBHs.
 They do not have extreme BH masses, $\rm H\beta$ FWHMs, 
 or optical luminosities either.
 However, we caution that the estimated BH masses and 
 accretion rates may have large uncertainties
 (see Section~\ref{sec:intro}). 
These results and the $\approx(15\textrm{--}24)\%$ 
occurrence rate of extreme \xray\ variability among SEAMBHs 
suggest that, although the high accretion rate may be a key factor 
for the extreme \xray\ variability in AGNs, 
 there should be some factor, other than the BH mass, $\rm H\beta$ 
 FWHM, or optical luminosity, that also influences the \xray\ 
 variability.

\subsection{Possible Scenarios for Extreme \xray\ Variability}
\label{sec:model}
The extreme \xray\ variability of \hbox{SDSS~$\rm J0751+2914$}
and other similar AGNs is unlikely the intrinsic variability of 
typical AGNs which rarely exceeds a variability factor of
200\% (see Section~\ref{sec:intro}). There are probably physical
causes for the observed extreme phenomenon. 
Here we discuss two popular scenarios, 
the reflection and partial covering absorption models,  
frequently adopted to explain the extreme \xray\ variability 
of AGNs.

The reflection model proposed by \cite{Ross2005} has been
 applied to explain the \xray\ properties of many extremely
 variable NLS1s (e.g.,
 \hbox{1H~$0707\textrm{--}495$}: \citealt{Fabian2004,Fabian2012};
 Mrk~335: \citealt{Grupe2007a,Grupe2008a,Gallo2015}; 
 NGC~4051: \citealt{Ponti2006}; 
 IRAS~$13224-3809$: \citealt{Ponti2010,Jiang2018})
 and luminous quasars (e.g., PG~$0844+349$: \citealt{Gallo2011};
 PHL~1092: \citealt{Miniutti2012}).
It proposes that in addition to the continuum emission observed
 directly, part of the primary \xray\ emission from the corona
 is reflected to the line of sight by the accretion disk.
The observed \xray\ variability is the result of changes in the
 height of the corona (simplified as a point source above the BH 
 in the "lamppost" geometry; e.g., 
 \citealt{Fabian2003,Miniutti2003}).
As the \xray\ point source approaches the BH,
the gravitational light bending gets stronger and it 
reduces the number of \xray\ photons reaching the observer
 \citep{Miniutti2004}.
 More primary \xray\ power-law photons are reflected by the 
 inner disk, so that the ratio of the reflection flux
 to the observed power-law continuum flux is larger.
In the low state, the $\rm <10~keV$ spectrum is dominated 
by a soft-excess component and a broad Fe K$\alpha$ 
 emission line at $\sim 6\textrm{--}7~$keV, 
with a flat spectral shape and curvature emerging 
in the $\rm \approx2\textrm{--}6~keV$ band.  
The observed \xray\ and multiwavelength properties of 
\hbox{SDSS~$\rm J0751+2914$} are generally consistent with 
this scenario, although the \hbox{low-state} spectrum is likely dominated by the soft excess component
and we did not observe the broad Fe K$\alpha$  
 emission due to the limited photon counts.

Although the reflection model usually describes well the spectra
 of extremely variable AGNs, one main caveat is that the disk
 thickness is assumed to be very small and negligible compared to
 the height of the corona. However, the vertical structure of the
 accretion disk should not be neglected for AGNs with 
 $\lambda_{\rm Edd}\ga 0.1$, at which the accretion flow may
 become advection dominated in the radial direction, and the 
 inner disk becomes geometrically thick \citep[e.g.,][]
 {Abramowicz1980,Abramowicz1988,Wang2003,Ohsuga2011,
 Wang2014b, Jiang2016,Jiang2017}.
Such a thick disk may produce a reflected spectrum 
 different from that from a thin disk, and it is also likely
 to obscure the \hbox{X-rays} from the corona when the
 inclination angle is large
 (e.g., \citealt{Luo2015,Ni2018,Taylor2018}; and references
 therein). Therefore, for extremely \xray\ variable AGNs that
 typically have high accretion rates ($\lambda_{\rm Edd}\ga 0.1$), 
 the geometry of the accretion disk probably needs to  
 be considered in the modeling of the \xray\ variability.
 
The partial covering absorption scenario 
\citep[e.g.,][]{Tanaka2004,Turner2009b,Miniutti2012}
depicts that the observed \xray\ variability
is attributed to the variation of the covering factor,
ionization, and column density of the absorber.
Under this scenario, 
 the unabsorbed part of the primary \xray\ emission dominates 
 the $\la2$~keV soft \hbox{X-rays}, and the absorbed component 
 dominates the hard \hbox{X-rays}. This model can also explain the
  \xray\ variability of \hbox{SDSS~$\rm J0751+2914$}.
The origin and physics of the partial covering absorption material
are not well understood. 
 Since there is no coordinated UV/optical variability in general,
 the partial covering absorber must be located within 
the BLR and close to the BH, otherwise it may 
absorb UV/optical photons.  

\subsection{Partial Covering Absorption by A Thick Disk/Outflow}
\label{subsec:thickdisk}
Since the extreme \xray\ variability of AGNs appears to be 
associated with high accretion rates (Section~\ref{subsec:sample_ex}),
we consider that a geometrically thick accretion disk and its
associated dense outflow may serve as the partial covering absorber
for blocking the central \xray\ emission.
For AGNs with high accretion rates, the inner accretion disk is
 expected to be geometrically thick, and such an accretion disk is
  likely to produce a strong outflowing wind 
 \citep[e.g.,][]{Ohsuga2011,Takeuchi2014,Jiang2016,Jiang2017}.
The thick disk and its associated outflow can absorb partially the 
\xray\ emission when the inclination angle is large 
(see Figure~18 in \citealt{Luo2015} and Figure~1 
 in \citealt{Ni2018}). 
As the height/size of the corona 
changes like in the reflection scenario, the covering factor of the 
thick disk/outflow with respect to the \xray\ corona changes
 accordingly, resulting in the observed \xray\ variability. 
The thick disk/outflow does not affect the observed UV/optical 
continuum or emission lines. This model can explain the \xray\ and 
multiwavelength properties of \hbox{SDSS~$\rm J0751+2914$} and 
other AGNs with extreme \xray\ variability.
It also naturally explains the small occurrence rate
 of extremely \xray\ variable AGNs with high accretion rates, 
 as only AGNs with a line of 
 sight close to the edge of the thick disk/outflow may experience 
 variable partial covering \xray\ absorption when 
 the corona height/size changes.
We note that if the absorber is the outflow associated with the inner 
disk, it must be relatively compact and cling to the disk;
otherwise, the fraction of extremely \xray\ variable AGNs would be 
much larger.

\subsubsection{Connections to Weak Emission-Line Quasars}
Our proposed scenario above shares the same basic 
nature as that for weak 
emission-line quasars (WLQs) in \cite{Luo2015},
and WLQs are generally considered to have high accretion rates 
\citep[e.g.,][]{Luo2015,Ni2018,Marlar2018}.
Therefore, WLQs and 
high accretion-rate AGNs with extreme
\xray\ variability are probably connected.
As proposed by \cite{Luo2015},
the inner puffed-up disk in a rapidly accreting
WLQ could block the nuclear ionizing emission from reaching the
the BLR, which results in the observed weak high-ionization UV emission
lines (e.g., \ion{C}{4}).
We investigated the UV emission lines of the extremely
\xray\ variable NLS1s and quasars listed in
Table~\ref{tbl-sample}.
PHL~1092 exhibits a weak, blueshifted, and asymmetric \ion{C}{4}
 emission line similar to those in WLQs \citep[e.g.,][]{Miniutti2012}.
Two NLS1s, IRAS~$13224-3809$ and 1H~$0707-495$, also show
weak \ion{C}{4} emission lines with equivalent widths (EW) less
 than $15~{\textup{\AA}}$ \citep{Leighly2004},
satisfying the \ion{C}{4} EW criterion for WLQs
\citep{Ni2018}. The other three objects (PG~$1211+143$,
 PG~$0844+349$, and Mrk~335) do not show weak \ion{C}{4} lines
  \cite[e.g.,][]{Baskin2005,Wu2009,Tang2012}. 
 As the emission-line strength is influenced by many factors, 
 including anisotropic continuum and line emission, gas metallicity, 
 and BLR geometry \citep[e.g.,][and references therein]{Luo2015},
 it is probably not surprising to observe typical \ion{C}{4} line 
 strengths in a significant fraction of high accretion rate AGNs with 
 thick inner accretion disks.
 A UV spectrum is needed to check if \hbox{SDSS~$\rm J0751+2914$}
has a weak \ion{C}{4} emission line similar to those of PHL~1092 and
WLQs.

Considering that WLQs and AGNs with extreme
\xray\ variability likely share the same nature, 
we expect that some WLQs with large inclination
 angles would also vary extremely in \hbox{X-rays}. 
 Specifically, all the WLQs that were observed to be extremely \xray\
 weak could be extremely \xray\ variable. Unfortunately, most of
the extremely \xray\ weak WLQs have only been observed once and 
we cannot assess their variability. Moreover, 
if the variability timescale scales with the BH
mass, it would take a much longer time to detect \xray\ variability 
in WLQs with BH masses that are typically one order of magnitude 
larger than those of the extremely \xray\ variable quasars listed in 
Table~\ref{tbl-sample}.
Among the 32 WLQs in the representative sample of \cite{Ni2018}, 
there are two extremely \xray\ weak ($f_{\rm weak}>10$, 
$\Delta\alpha_{\rm OX}<-0.384$) WLQs plus 10 
\xray\ undetected WLQs that could also be extremely \xray\ weak. 
The fraction ($2/32\textrm{--}12/32$) is in general consistent 
with the fraction of extremely \xray\
variable AGNs among high accretion rate AGNs, supporting a common
origin for these extreme phenomena.

\section{summary and future work}\label{sec:sum}
In this paper,
we report the discovery of extreme \xray\ variability in a 
type 1 quasar: SDSS~J$0751+2914$.
It is powered by a super-Eddington accreting BH with a mass of 
$\sim 1.6\times 10^7~M_\odot$.
Based on archival observations, we have constrained its
 \xray\ and UV/optical properties in different epochs.
SDSS~J$0751+2914$ shows extreme \xray\ variability by a factor of 
 larger than 10, and it lacks significant UV/optical 
 variability coordinated with its \xray\ variability. 
We also investigated other extremely \xray\ variable AGNs
with similar properties in the literature.
Our main results are as follows:

\begin{enumerate}
\item
In general,
a single power-law model modified by Galactic absorption describes
well the \hbox{0.3--10} keV spectra of the three \xray\
 observations.
The spectral fitting yielded three steep power-law photon indices,
$\rm \Gamma_s=2.89^{+0.14}_{-0.13}$, $3.07^{+0.37}_{-0.37}$,
and $3.06^{+0.18}_{-0.17}$, for the high, intermediate,
and low \xray\ flux states, respectively.
See Section~\ref{subsec:xspec}.

\item
Between 2013 May 27 and 2015 May 4 the observed
0.5--2 keV flux of \hbox{SDSS~$\rm J0751+2914$}
dropped by a factor of $21.6\pm 2.4$.
Since its UV flux shows little change in this period,
it became extremely \xray\ weak in 2015 May
with a steep X-ray-to-optical power-law slope
($\alpha_{\rm OX}$) of $-1.97$, corresponding to an
\xray\ weakness factor of $36.2$ at rest-frame 2~keV. 
See Sections~\ref{subsec:xvar} and \ref{subsec:uvvar}.

 \item
The optical continuum and emission lines of
\hbox{SDSS J0751+2914} show little change between
the high and low \xray\ flux states, which indicates
an almost constant accretion rate. 
See Section~\ref{subsec:opt}.

\item
Most of the extremely \xray\ variable 
  AGNs reported in the literature are NLS1s and NL type 1 quasars that
  have high accretion rates. But only a small fraction of such objects
  are extremely \xray\ variable by factors of more than 10.
The fractions of extremely variable \xray\ sources among NLS1s
 \citep{Grupe2010} and SEAMBHs
\citep{Du2014,Du2015,Du2018}
are estimated to be $\approx 8\textrm{--}11\%$ and $\approx 15\textrm{--}24\%$, respectively. 
See Sections~\ref{subsec:sample_ex} and \ref{subsec:frac}

\item
We reviewed the reflection and partial covering 
absorption models, frequently applied to explain the extreme
\xray\ variability of NLS1s. 
Either model can explain the overall observational data  
for SDSS~J$0751+2914$.
We further propose that a thick accretion disk and its 
associated 
outflow can serve as the absorber in the partial covering 
absorption scenario.
This model can explain the \xray\ and multiwavelength 
properties of SDSS~J$0751+2914$ and other AGNs with extreme 
\xray\ variability. It also explains naturally the small 
fraction of extremely \xray\ variable AGNs among AGNs with 
high accretion rates, as only AGNs with a
line of sight close to the edge of the thick disk/outflow may 
experience variable partial covering \xray\ 
absorption when the height/size of the corona changes.
We also discuss the connections between extremely X-ray 
variable AGNs and WLQs. 
See Sections~\ref{sec:model} and \ref{subsec:thickdisk}.
\end{enumerate}

We tried to piece together a complete picture of the population of 
extremely \xray\ variable AGNs by exploring the archival data and 
results for SDSS~J$0751+2914$ and other NLS1s and NL type 1 quasars 
showing similar properties. However, previous observations and studies
were mainly focused on individual objects, and there are
a lot of uncertainties when we tried to understand the nature of these 
objects as a unified population. For example, the duty cycle of the 
extremely \xray\ weak state and the occurrence rate of extreme \xray\
 variability among AGNs with high accretion rates
that we estimated in the current paper have substantial uncertainties, 
which affect our interpretation of the physical nature. 
We consider several possible future efforts below
that may help constrain better the properties of SDSS~J$0751+2914$ 
and other similar AGNs, which would ultimately help us understand
better the central engine of accreting BHs.

Multi-epoch monitoring observations with \xmm\ or \chandra\ are 
required to obtain a longer term \xray\ light curve 
of SDSS~J$0751+2914$, for the purpose of constraining its 
duty cycle of the extremely \xray\ weak state. Also, 
considering that its optical flux has significantly increased 
in the 2016--2017 RM monitoring period, it is of interest to
 investigate its current \xray\ state.
  In addition,
we estimate that a deeper \xmm\ observation on 
\hbox{SDSS J$0751+2914$} with 
$\approx 70$~ks exposure time could reveal 
 a flat/hard $>2$~keV \xray\ spectrum similar to the other extremely
 \xray\ variable AGNs (see Section~\ref{subsec:sample_ex}), if it is
 still in the extremely \xray\ weak ($f_{\rm weak} \sim 36$) state. 
  A \nustar\ observation is also required to examine if there is 
  a hump in the $\approx 20\textrm{--}30~\rm keV$ spectrum  
  (see Section~\ref{subsec:sample_ex}).
  
 We proposed a connection between extremely \xray\ variable
AGNs and WLQs. A UV spectrum of SDSS J$0751+2914$ will be able
to determine whether it has any weak UV emission lines 
(e.g., \ion{C}{4}).
Moreover, a UV spectroscopic survey of extremely \xray\ variable
AGNs (including exploring archival data) will allow us to examine such
a connection systematically.  A UV spectrum of SDSS J$0751+2914$ will
also help us to rule out the possibility that it is a BAL quasar.

We investigated the \xray\ variability of the NLS1s in 
 \cite{Grupe2010}, and we adopted their \xray\ analysis results to 
 estimate the occurrence rate of extreme \xray\ variable NLS1s.
New archival observations have become available for some of these NLS1s
 since \cite{Grupe2010}. These data will help to constrain the 
 fraction of extremely \xray\ variable AGNs with greater certainty.
 In addition, a systematic multi-epoch \xray\ survey on
the SEAMBHs is required to discover more extremely variable \xray\
 sources and constrain better the occurrence rate of extreme 
 \xray\ variability among this population.

~\\

We thank the referee for the helpful comments and suggestions.
We thank Yanmei~Chen, Qiusheng~Gu, and Zhiyuan~Li
for helpful discussions.
We acknowledge financial support from
the National Natural Science Foundation of China
grant 11673010 (H.L., B.L.),
National Key R\&D Program of China grant 2016YFA0400702 (H.L., B.L.),
National Thousand Young Talents program of China (B.L.).
W.N.B. acknowledges financial support from NASA ADP grant
  80NSSC18K0878 and CXC grant GO6-17083X.

\bibliographystyle{aasjournal}
\bibliography{ms}

\begin{turnpage}

\begin{deluxetable*}{lcccccccccc}
\tablewidth{0pt}
\tablecaption{ AGNs with Extreme \xray\ Variability
}
\tablehead{
\colhead{Object} &
\colhead{$z$}   &
\colhead{log($M_{\rm BH}/M_\odot$)} &
\colhead{Note \tablenotemark{a}} &
\colhead{FWHM(H$\rm \beta$)}  &
\colhead{$L_{5100~\textup{\AA}}$} &
\colhead{log~$L_{\rm Bol}$} &
\colhead{log~$\rm\dot{\mathscr{M}}$}&
\colhead{$\lambda_{\rm Edd}$} &
\colhead{References} \\
\colhead{ }  &
\colhead{ }  &
\colhead{  } &
\colhead{ }  &
\colhead{$\rm km~s^{-1}$}  &
\colhead{$\rm erg~s^{-1}$ } &
\colhead{$\rm erg~s^{-1}$ }  &
\colhead{ }  &
\colhead{ }  &
\colhead{ }
}
\startdata
\multicolumn{11}{c}{Quasars} \\
 \midrule
SDSS $\rm J0751+2914$& 0.121  & 7.20 & M & 1679 &44.21  &45.22&1.45 &0.7 & \cite{Du2018} \\
PG $1211+143$        & 0.085  & 7.87 & M & 2012 &44.73  & 45.72&0.84 &  0.5 & \cite{Du2015,Danehkar2018} \\
PHL 1092             & 0.396  & 8.48 & X & 1800 &45.45  &46.65&0.65 &1.11 &\cite{Dasgupta2004,Nikolajuk2009,Miniutti2012} \\
PG $0844+349$        & 0.064  & 7.66 & M & 2694 &44.22  & 45.4 &0.50 &0.36 &\cite{Vasudevan2009}, \cite{Du2015}\\
 \midrule
 \multicolumn{11}{c}{Selected Lower-luminosity Counterparts} \\
 \midrule
1H $0707-495$          & 0.0411 & 6.60 & R   & 980 & 43.52 & 44.47&1.57 &0.6 &  \cite{Done2016}\\
IRAS $13224-3809$\tablenotemark{b}     & 0.0667 & 6--7 & -- & 650 & 43.94 &44.53 &--&$0.3\textrm{--}3$ &\cite{Jiang2018}\\
Mrk~335              & 0.0258 & 6.93& M & 1707 & 43.76 & 44.73 &1.27  & 0.42  &\cite{Du2015} \\
NGC~4051	& 0.00234& 5.42 & M&851& 41.96& 42.91 &1.59 & 0.21 &\cite{Du2015} 
\enddata
\tablecomments{
\tablenotetext{a}{Method for estimate of BH mass. M: reverberation mapping method; X: \xray\ excess variance method; R: scaling relation of \cite{Vestergaard2006}.}
\tablenotetext{b}{The BH-mass estimates of IRAS~$13224-3809$ in
 previous studies have large uncertainties. Listed in this table are 
 the possible ranges of estimated BH mass and Eddington ratio.
Its $L_{5100~\textup{\AA}}$ value was derived from an interpolation 
of the $B$- and $V$-band photometric data \citep{Ojha2009}.
 }
}
\label{tbl-sample}
\end{deluxetable*}
\end{turnpage}

\end{document}